\documentclass[10pt]{article}
\usepackage{amsmath}
\usepackage{amsfonts}
\usepackage{amssymb}
\usepackage{graphicx}
\usepackage{caption}
\usepackage{subcaption}
\usepackage{rotating}
\usepackage{makecell}
\usepackage{hyperref}
\usepackage{multirow}
\usepackage{lipsum}
\usepackage{calc}
\usepackage{float}
\usepackage{booktabs}
\usepackage{verbatim}

\usepackage{microtype}
\usepackage{tabularx}
\usepackage{xurl}
\usepackage{geometry}
\usepackage{authblk}
\PassOptionsToPackage{hyphens}{url}\usepackage{hyperref}

\usepackage[square,numbers]{natbib}

\newcolumntype{R}[1]{>{\raggedleft\let\newline\\\arraybackslash\hspace{0pt}}m{#1}}

\title{The Role of Search Engines in the Amplification and Suppression of LGBTIQ+ Polarization}
\author[1]{Ronja R{\"o}nnback}
\author[1]{Chris Emmery}
\author[1]{Marie \v{S}af\'{a}\v{r} Postma}
\author[2]{Filip Milde}
\author[3]{Jan Charv\'at}
\author[1,*]{Henry Brighton}
\affil[1]{Tilburg School of Humanities and Digital Sciences, Tilburg University, Tilburg, The Netherlands}
\affil[2]{We Are Fair, Prague, Czech Republic}
\affil[3]{Department of Political Science, Charles University, Prague, Czech Republic}
\affil[*]{Corresponding author: \href{mailto:h.j.brighton@tilburguniversity.edu}{\nolinkurl{h.j.brighton@tilburguniversity.edu}}}
\date{}

\begin{document}

\maketitle

\begin{abstract}
Search engines are used and trusted by hundreds of millions of people every day. However, the algorithms used by search engines to index, filter, and rank web content are inherently biased, and will necessarily prefer some views and opinions at the expense of others. In this article, we examine how these algorithmic biases amplify and suppress polarizing content. Polarization refers to a shift toward and the acceptance of ideological extremes. In Europe, polarizing content in relation to LGBTIQ+ issues has been a feature of various ideological and political conflicts. Although past research has focused on the role of social media in polarization, the role of search engines in this process is little understood. Here, we report on a large-scale study of 1.5 million search results responding to neutral and negative queries relating to LGBTIQ+ issues. Focusing on the UK, Germany, and France, our analysis shows that the choice of search engine is the key determinant of exposure to polarizing content, followed by the polarity of the query. Location and language, on the other hand, have a comparatively minor effect. Consequently, our findings provide quantitative insights into how differences between search engine technologies, rather than the opinions, language, and location of web users, have the greatest impact on the exposure of web users to polarizing Web content. 
\end{abstract}

\section{Introduction}

Search engines have the capacity to influence people’s preferences, opinions, and decisions on an immense scale. In 2023, for example, Bing reported more than 100 million active daily users despite only achieving a single-digit market share \cite{bing_2023}. This implies that Google, which has a market share of over 90\% in many parts of the world, may attract over 1 billion daily users \cite{statcounter_2024}. 
The scale of search engine use, along with findings that show that people often trust search engines more than social and traditional media \cite{edelman2024,edelman2025}, has fueled the concern that search engine users are susceptible to influence and manipulation \cite{epstein_2018_manipulating,aslett_online_2024}. 
For example, subtle changes to the ranking of politicized search results can shift the voting preferences of undecided voters from one candidate to another \cite{SEME}. Given this capacity to influence web users at scale, it is crucial to understand the ways in which search engines are biased toward certain views and opinions at the expense of others, and whether these biases undermine the integrity of the Web \cite{Introna_2000_shaping,wallace2017}. 

A key barrier to studying search engines is that, like many online platforms, their algorithms are hidden from public view \cite{rieder_2020_towards}. Moreover, countermeasures designed to block automated interactions make it technically challenging to conduct large-scale scientific analyses. 
While a number of large-scale monitoring studies have sidestepped this problem, they have tended to focus on the political bias of English language search results, considered only users located in the US, and restricted their attention to Google and occasionally Bing \cite{robertson_auditing_personalization_2018,robertson_auditing_bias_2018,Metaxa_2019,robertson_2023_users}. Building on this work, we set out to examine the role played by four independent search engines in the propagation and suppression of polarizing content in Europe. Polarization, which refers to a shift toward and the acceptance of ideological extremes \cite{dimaggio_1996_have}, is seen as a threat to the online information space by key European institutions \cite{fletcher_EU_Paliament_polarisation_2019,EC_lgbtq}. Polarizing content typically lacks credibility, tends to be misinformative, is frequently politicized, and often targets minority groups such as LGBTIQ+ people, which includes
people who are lesbian, gay, bisexual, transgender, intersex, and queer \cite{strand2021disinformation,graff_gender_and_global_right_2019,kovats_gender_glue_2015,edenborg_traditional_2023,grabowska-moroz_reframing_2021}. 

In this article, we examine to what extent search engines return low credibility, politicized results likely to fuel polarizing, extremist views toward LGBTIQ+ people. 
Taking a data-driven approach, we monitored four independent search engines (Google, Bing, Yandex, and Mojeek) over a 3-month period and collected over 1.5 million search results. By controlling the location from which queries were issued, the language they were issued in, and whether the queries conveyed neutral or negative perspectives on LGBTIQ+ issues, we show that the choice of search engine has the greatest impact on exposure to low credibility, polarizing content. Yandex and Mojeek, for example, return over eight times as many low credibility results as Google and Bing. Our findings provide quantitative insights into how differences between search engine technologies, rather than the views, opinions, language, and location of web users, have the greatest impact on the exposure of web users to polarizing Web content.

\section{Polarization, LGBTIQ+ rights, and search engine monitoring}\label{sec:lgbt_related_work}

Recent years have seen an increase in the polarization of traditional and social media \cite{kubin_role_SoMe_polarisation_2021,boxell_cross-country_polarisation,chinn_politicization_climate_2020}. 
This shift toward ideological extremes has negative consequences, such as threatening the functioning of democratic institutions \cite{ecker2024misinformation}, and undermining trust and adherence to the rule of law \cite{keymolen_DoorsOfJanus}. Although a range of social and political factors drive the process of polarization, the role of technologies such as social media platforms are increasingly under scrutiny \cite{agent-modelling_polarisation_2024,weismueller_falsehood_partisanship_for_polarisation_2024}. The questions of how and to what extent online platforms play a role in polarization have been posed in relation to several countries \cite{budak_misunderstanding_2024}, but beyond the observation that different countries appear to be affected to different degrees, much remains uncertain \cite{kubin_role_SoMe_polarisation_2021,fletcher_EU_Paliament_polarisation_2019,boxell_cross-country_polarisation,urman_political_polarisation_twitter}. In this article, we address the role of search engines in the amplification and suppression of polarizing content relating to LGBTIQ+ people. LGBTIQ+ rights are of particular interest due to a growing body of research identifying a connection between anti-gender ideology (which seeks to oppose women's rights as well as LGBTIQ+ rights \cite{strand2021disinformation,graff_gender_and_global_right_2019,norocel_disarticulation_2023}) and democratic backsliding, where democratic processes are undermined in favor of autocratic rule \cite{flores2023democratic,lombardo_gender_rights_litmus_test,grabowska-moroz_reframing_2021}. 

Because the character of online anti-LGBTIQ+ discourse varies depending on the language and the country where it occurs, we developed a multilingual dataset of search engine queries that we use to monitor search engines in different countries. For example, in Germany, polarizing discourse on Twitter/X often invokes the idea of vaccine-induced homosexuality \cite{locatelli_cross-lingual_2023}. In contrast, in France, it is more common to use LGBTIQ+ slurs to insult political figures and athletes, and invoke disinformation narratives about homosexuality and disease \cite{locatelli_cross-lingual_2023}. Focusing on these cultural differences, our investigation of the amplification and suppression of polarizing content addresses a perceived threat to the online environment recently highlighted by key European institutions \cite{fletcher_EU_Paliament_polarisation_2019,EC_lgbtq,strand2021disinformation}. 
In contrast to our data-driven focus on search engines, previous work has tended to explore the issue of LGBTIQ+ polarization from a social, legal, and security perspective. This work often relies on case studies and aims to provide a detailed examination of specific regions \cite{norocel_disarticulation_2023,geguchadze2021lgbtq,unal_Turkey_2024,Nato_russian_influence_in_baltic,edenborg_traditional_2023}, general narratives and strategies \cite{strand2021disinformation,jalonen_identity_2023,edenborg_traditional_2023,corrales_homophobic_2022,datta_tip_of_iceberg_antilgbt_funding}, and potential geopolitical motivations for the polarization of LGBTIQ+ issues \cite{norocel_disarticulation_2023,paternotte_disentangling_2018,karlsen_divide_2019,jalonen_identity_2023,edenborg_traditional_2023,datta_tip_of_iceberg_antilgbt_funding}. Our approach complements this body of work, and illustrates how its limitations can be addressed by large-scale algorithmic monitoring \cite{haim_recommends-agent-based_2020}.

\subsection{Search engines and search engine monitoring} \label{sec:se_overview}

Google has a search engine market share of over 90\% in many parts of the world \cite{statcounter_2025_world}. The main exceptions are Baidu which has a 46\% share in China \cite{statcounter_2025_china}, and Yandex which has a 75\% share in Russia \cite{statcounter_2025_russia}. In Europe, the main competitors to Google are Bing and Yandex, which both have a market share of roughly 4\% \cite{statcounter_2025_europe}. Few independent alternatives exist although a number of search engine companies provide privacy-preserving interfaces to Google and Bing \cite{duckduckgoWhereDuckDuckGo}. A notable exception is Mojeek, which is an independent search engine based in the UK that prioritizes web content from less well-known and less authoritative sources \cite{mojeekIndependentResults}. Because one of our goals is to investigate variation among search engine technologies, we will focus on four independent search engines that serve Europe: Google, Bing, Yandex, and Mojeek. Our approach to search engine monitoring has four key features:

\begin{enumerate}

\item {\em Active Monitoring}. 
Several large-scale studies have used technologies such as browser plug-ins to intercept search results delivered to human users \cite{robertson_auditing_personalization_2018,robertson_auditing_bias_2018,robertson_2023_users}. The advantage of these approaches is that the search results being collected are responses to the queries of real users engaging with search engines. The disadvantage is that by passively monitoring search engines through the lens of human activity, there is no way of controlling which queries are issued, when, and from which location. In this study, we perform active monitoring by automating search engine interactions directly, and without any human activity driving the monitoring process. This allows us to design our monitoring study to address specific, experimentally controlled questions. 

\item {\em Multilingual Query Dataset.} Although several studies have monitored search engines by issuing a controlled series of queries \cite{Metaxa_2019,norocel_data_voids_2023}, they have tended to be small-scale studies or focus on a narrow range of queries \cite{urman_conspiracies_in_SEs,makhortykh_Google_Yandex_SmartVoting_2022,leon_search_US_election_conspiracy_2020_2024,kuznetsova_search_Ukraine_disinfo_2024,makhortykh_search_US_election_2025,urman_US_presidential_SEs_2022}. In this study, we used a large, multilingual, and structured dataset of 3766 queries covering a range of LGBTIQ+ issues. This dataset was compiled in collaboration with experts in the field of LGBTIQ+ rights (see Section~\ref{sec:queries}), and for each issue, we include queries conveying neutral and negative perspectives on LGBTIQ+ people. The entire batch of queries was issued independently from 32 regional monitoring hubs spanning eight countries. For each hub, all queries were issued using ethically sourced residential proxies with a registered location within the specified region (for example, to monitor from London we used residential proxies registered in London).

\item {\em Polarizing Content and Disinformation}. Recent search engine monitoring studies have tended to focus on the political bias of search results, particularly during critical periods of the US election cycle \cite{robertson_auditing_personalization_2018,robertson_auditing_bias_2018,Metaxa_2019,robertson_2023_users,makhortykh_search_US_election_2025,leon_search_US_election_conspiracy_2020_2024}. One barrier to addressing the issue of polarization is that there is no formal definition of the term. On the other hand, prior work has considered how politically partisan views are a key feature of polarizing content \cite{makhortykh_search_US_election_2025,garimella_polarisation_of_online_news_browsing_2021}, as well as the reliability and credibility of the source of the content \cite{weismueller_falsehood_partisanship_for_polarisation_2024}. We will integrate these two features of polarizing content by combining measures of political bias and content credibility. Consequently, we focus not on user polarization, but on potentially polarizing content that web users are exposed to \cite{budak_misunderstanding_2024}. Underlying our approach is the idea that different groups of web users can be exposed to different kinds of content, and this may 
exacerbate ideological disagreement and hostility \cite{rekker2022polarisation_and_facts}. To determine whether search results reference potentially polarizing web content, we use the NewsGuard dataset. NewsGuard is a commercial dataset that details the political orientation and journalistic credibility of approximately 12,000 web domains, as well as providing flags that indicate cases of disinformation (\cite{NewsGuard}, see Section~\ref{sec:newsguard}). 

\item {\em Analysis of Unpersonalized Organic Search Results}. When logged into a user account, Google, Bing, and Yandex deliver personalized search results (see, for example, \cite{GoogleSearchHelp_Personalization}). Mojeek, in contrast, does not collect user information and only delivers unpersonalized search results. In order to compare different search engines on an equal footing, we will sidestep the complexities of comparing different personalization algorithms by focusing on organic, unpersonalized search results. Similarly, in order to make direct comparisons, we will not consider page components such as news, videos, and other potentially search engine specific aspects of the search results. As such, our approach examines the basic ability of search engines to supply ranked search results. Although personalization is a key feature of the major search engines, several studies report that the differences between generic results and personalized search results tend to quite be small \cite{hannak_personalization_2013,personalisation_location,robertson_auditing_personalization_2018}.
\end{enumerate}

In summary, our approach to search engine monitoring is specifically tailored to the European context, and it allows us to study the relative influence of location, language, query composition, and search engine technology on the amplification and suppression of polarizing LGBTIQ+ content. 

\section{Data collection through active monitoring}

Between the 19th of April and the 11th of June 2024 we monitored Google, Bing, Yandex, and Mojeek in The United Kingdom, France, Germany, the Netherlands, Poland, Slovakia, Czech Republic, and Hungary. Within each country, the entire batch of queries for that country was issued independently, and from each of four separate regional hubs. This process was repeated three times, resulting in the collection of more than 1.5 million search results. We then analyzed these search results using the NewsGuard dataset, described below. Of the 8 countries we monitored, however, NewsGuard only provides ratings for English, German, and French language web domains. For this reason, in the following analysis we focus on results collected in four locations in the UK (London, Cardiff, Edinburgh, and Belfast), Germany (Berlin, Dresden, Frankfurt, and Munich), and France (Paris, Marseilles, Bordeaux, and Strasbourg). This subset of our results includes 561,317 search results (120,906 collected in France, 220,084 in Germany, and 220,327 in the UK). To further clarify this process, we first detail the query dataset, how search engine interactions were automated, and which aspects of the NewsGuard dataset we used in our analysis. 

\subsection{A multilingual query dataset}\label{sec:queries}

To systematically study the extent to which search engines expose users to LGBTIQ+ related web content, we created a structured, multilingual dataset of queries relating to LGBTIQ+ issues. For each language, we employed a local specialist on LGBTIQ+ rights to guide the construction of queries relating to 7 categories: general LGBTIQ+ issues, LGBTIQ+ disinformation narratives, gender issues, LGBTIQ+ relationships, LGBTIQ+-related personalities, and known anti-LGBTIQ+ extremists (for examples, see Table \ref{tab:brief_example_queries}; Table \ref{tab:base_conditions} in the Appendix provides a full description of each category and additional examples in English). All queries convey either a neutral perspective (e.g., ``gay rights"), or a negative perspective (``gay agenda"), which we refer to as the polarity of the query. An approximately equal number of neutral and negative queries are included in each category. A key feature of the query dataset is that it reflects significant cultural differences in the issues, terminology, and personalities that feature in discussions and conflicts surrounding LGBTIQ+ issues \cite{locatelli_cross-lingual_2023,jalonen_identity_2023,edenborg_traditional_2023}. As a result, it is rarely the case that a query in one language has a direct translation and has the same meaning and relevance in another language. For example, queries about LGBTIQ+-related personalities differ significantly from one language to another because the regional experts responsible for compiling the queries worked independently from each other, and often selected individuals who were only nationally relevant. In contrast, the category of anti-LGBTIQ+ extremists contains only minimal differences between languages because, for each language, the queries reference the same globally relevant set of individuals identified by the Southern Poverty Law Center \cite{SPLC_extremist_files}. In total, our analysis centers on 1,297 queries (288 in French, 528 in German, and 502 in English). 

When presenting our results, we will draw attention to a subtlety in the interpretation of query polarity. For the 5 query categories that cover LGBTIQ+ topics, the query polarity refers to whether the query conveys a neutral or negative perspective on LGBTIQ+ people. In contrast, for the 2 categories that cover queries about individuals (LGBTIQ+-related personalities and anti-LGBTIQ+ extremists), a negative polarity refers to a negative perspective on the individual in question, rather than a negative perspective on LGBTIQ+ people (e.g., ``Chaya Raichik is homophobic''). Similarly, a query conveying a neutral perspective on an individual means that the query is neutral with respect to that person, even though their views may be associated with anti-LGBTIQ+ sentiment (e.g. ``Chaya Raichik is right''). During our analysis, we will separate the analysis of queries about people and queries about topics to avoid any confusion. 

\begin{table}[ht]
    \caption{Eight example queries, their associated query category, their polarity, and language. All queries were issued in lower case.}
    \label{tab:brief_example_queries}
    \footnotesize
    \centering
    \begin{tabular}{p{0.3\textwidth} p{0.25\textwidth} p{0.16\textwidth} p{0.12\textwidth}}
    \toprule
    \textbf{Query} & \textbf{Query Category}   & \textbf{Query Polarity} & \textbf{Language} \\ [0.5ex] 
    \midrule
    homo diktatur     & General LGBTIQ+ issues  & Negative   & German      \\
    lgbt bewegung    & General LGBTIQ+ issues  & Neutral    & German      \\
    gays spread corona virus & LGBTIQ+ disinformation       & Negative   & English      \\
    gay people and stds  & LGBTIQ+ disinformation & Neutral    & English      \\
    gender nazis      & Gender issues      & Negative   & English      \\
    gender dysphoria     & Gender issues     & Neutral    & English      \\
    menace sur la famille par les homosexuels   & LGBTIQ+ relationships  & Negative   & French      \\
    gpa et lgbt    & LGBTIQ+  relationships  & Neutral    & French      \\
    \bottomrule
    \end{tabular}
\end{table}

\subsection{Automated search engine interactions}

We define a search engine interaction as the process of navigating to the homepage of the search engine, issuing a query, and retrieving the first ten ranked search results. In our study, all search engine interactions were automated, and all available localization options were selected for each search engine during these interactions (see Appendix \ref{app:localization}). These localization options range from selecting different services with different URLs in the case of Google (e.g., {\sf Google.co.uk} vs {\sf Google.de}), to different settings that restrict results to certain regions and languages in the case of Bing and Mojeek. Queries were always issued in the local language, and all internet traffic between our monitoring systems and the search engines was routed through ethically-sourced residential proxies located in the respective countries and regions. To allow direct comparisons between search engines, we will focus on ranked search results appearing in the main results panel. The structure of information boxes appearing in side panels, as well as separate panels focusing on news, images, and videos, often differs from one search engine to another, so we chose not to consider these additional information sources in our analysis. All search engine interactions were automated using the Chromium browser installed on the Linux operating system, and each interaction was automated using a clean browser session containing no stored cookies and no information relating to previous browser use. To minimize the possibility of personalization, no user accounts were used or associated with past use of the browser. 

\subsection{Domain-level political orientation and credibility ratings}\label{sec:newsguard}

\begin{table}[t]
    \centering
    \caption{The four features of the NewsGuard reliability dataset used in our analysis. For each web domain, we consider the credibility score, political orientation, topics covered, and disinformation flags. In Appendix \ref{app:ng_coverage} we detail the degree to which the search results we have collected are documented by the NewsGuard dataset. }
    \small
    \begin{tabular}{l p{0.7\textwidth}}
    \toprule
    \textbf{NewsGuard feature}     & \textbf{Description}   \\ 
    \midrule
    Credibility score & A score ranging from 0 to 100 that indicates how credible the web domain is. A score of 100 indicates that the website complies with all of NewsGuard's criteria of journalistic integrity, and is highly credible. A domain with a score below 60 is considered as untrustworthy by NewsGuard. Not all web domains are receive a score. For example, large platforms (e.g., Facebook) and domains associated with satire are not assigned a score.  \\[12mm] 
    Political orientation & The political orientation of the web domain, which is categorized as being either left-leaning, right-leaning, or having no political bias. \\[5mm]
    Topics covered & A list of the general topics covered by a web domain. For example, topics include political news or commentary, personal finance, health, religion, or conspiracy theories.   \\[8mm] 
    Disinformation flags & For each web domain, a list of flags that detail examples of misinformation or disinformation. Each flag refers to a category such as health information, the Ukraine conflict, or election integrity. \\
    \bottomrule
    \end{tabular}
    \label{tab:newsguard_features}
\end{table}

To quantify the extent to which search results reference low-credibility and potentially polarizing web content, a reliable way of characterizing web content is required. For large-scale studies referencing millions of URLs, reliably determining these characteristics at the level of the webpage is infeasible \cite{ronnback2025automatic}. Instead, previous studies have inferred document-level characteristics from properties of the web domains on which the documents were hosted \cite{Metaxa_2019,robertson_2023_users,robertson_auditing_personalization_2018}. For example, content appearing on \url{cnn.com} is assumed to be left-leaning, whereas content on \url{foxnews.com} is assumed to be right-leaning. While such inferences may not be valid for platforms that host varied content (such as YouTube, Facebook, and Twitter/X), for many web domains, inferences about specific pages can accurately be made from knowledge of the journalistic practices associated with the domain. In the following analysis, we use domain characteristics provided by a commercially available dataset developed by NewsGuard Technologies \cite{NewsGuard}. NewsGuard uses a systematic and transparent methodology to characterize the journalistic practices, credibility, and political orientation of over 10,000 web domains. Specifically, our analysis considers four features that are used to characterize web domains: credibility score, political orientation, associated topics, and disinformation flags (see Table \ref{tab:newsguard_features}). The credibility score ranges from 0 to 100, and is a weighted sum of nine component scores that cover various aspects of credibility and transparency, such as whether the web domain frequently publishes deceptive headlines. The political orientation of a domain is labeled as either left-leaning, right-leaning, or non-political. Associated topics are provided as a list of categories that describes the content of domain, such as personal finance, religion, or, most usefully for our purposes, conspiracy theories. Finally, the disinformation flags associated with a domain is a list of topics that this domain has published disinformation on. For reasons discussed above, all domains associated with platforms are excluded from our analysis. Many, but not all, domains referenced in our search results are covered by NewsGuard. We detail these coverage statistics per search engine and country in Appendix \ref{app:ng_coverage}.

\section{Results} \label{sec:results}

Recall that our goal is to understand the extent to which search engines return results that reference politically partisan, low-credibility content that is likely to fuel polarization. Specifically, we will examine the relative contribution of location, query category, query polarity, and the choice of search engine.

\subsection{Differences between countries}

\begin{figure}
    \centering
    \begin{subfigure}[t]{0.49\textwidth}
        \centering
        \includegraphics[height=3.8cm]{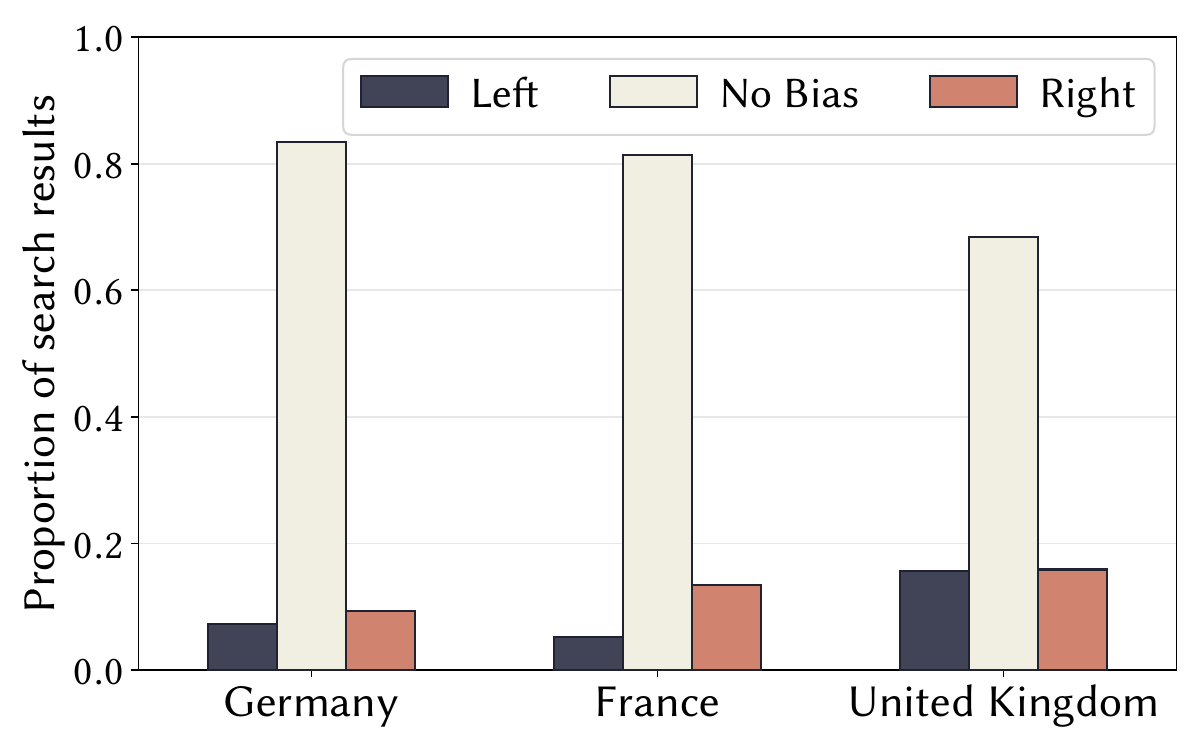}
        \caption{Political orientation for queries about topics.}
        \label{fig:political_lean_per_country_general}
    \end{subfigure}%
    \begin{subfigure}[t]{0.49\textwidth}
        \centering
        \includegraphics[height=3.8cm, trim=32 0 0 0, clip]{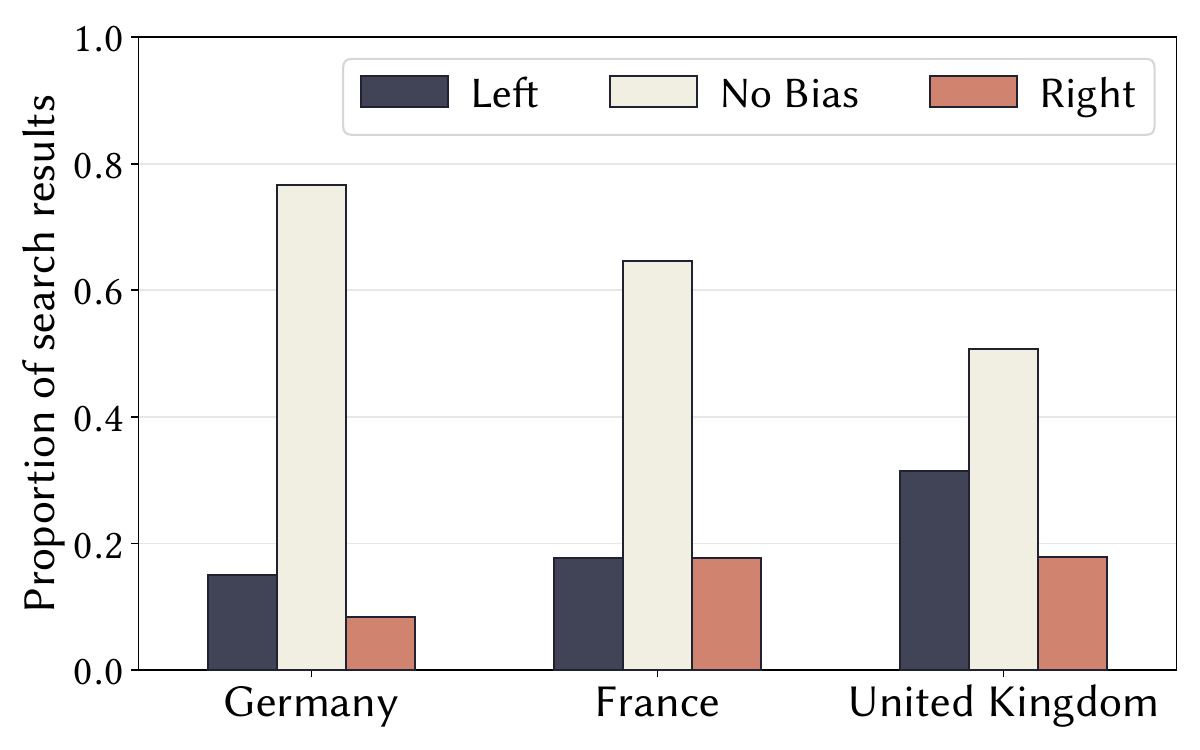}
        \caption{Political orientation for queries about individuals.}
        \label{fig:political_lean_per_country_people}
        \hspace{0.5cm}
    \end{subfigure}
    \begin{subfigure}[t]{0.49\textwidth}
        \centering
        \includegraphics[height=3.8cm]{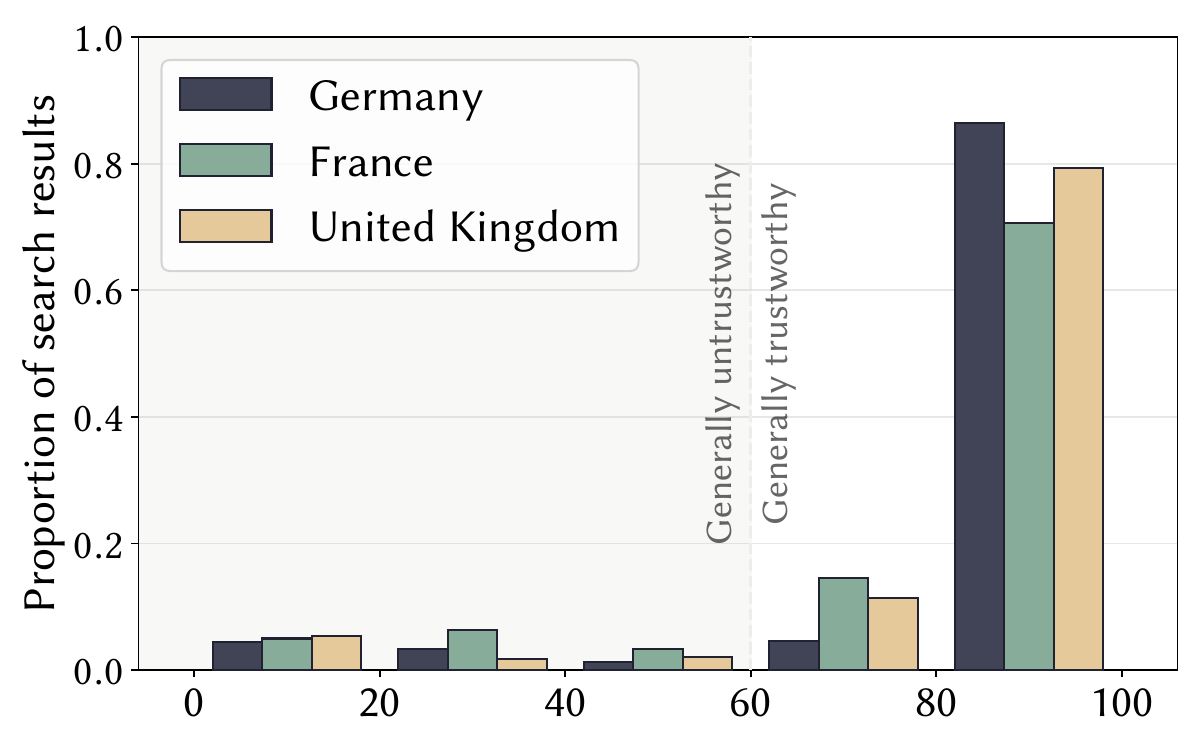}
        \caption{Credibility scores for queries about topics.}
        \label{fig:credibility_distribution_per_country_general}
    \end{subfigure}
    \begin{subfigure}[t]{0.49\textwidth}
        \centering
        \includegraphics[height=3.8cm, trim=32 0 0 0, clip]{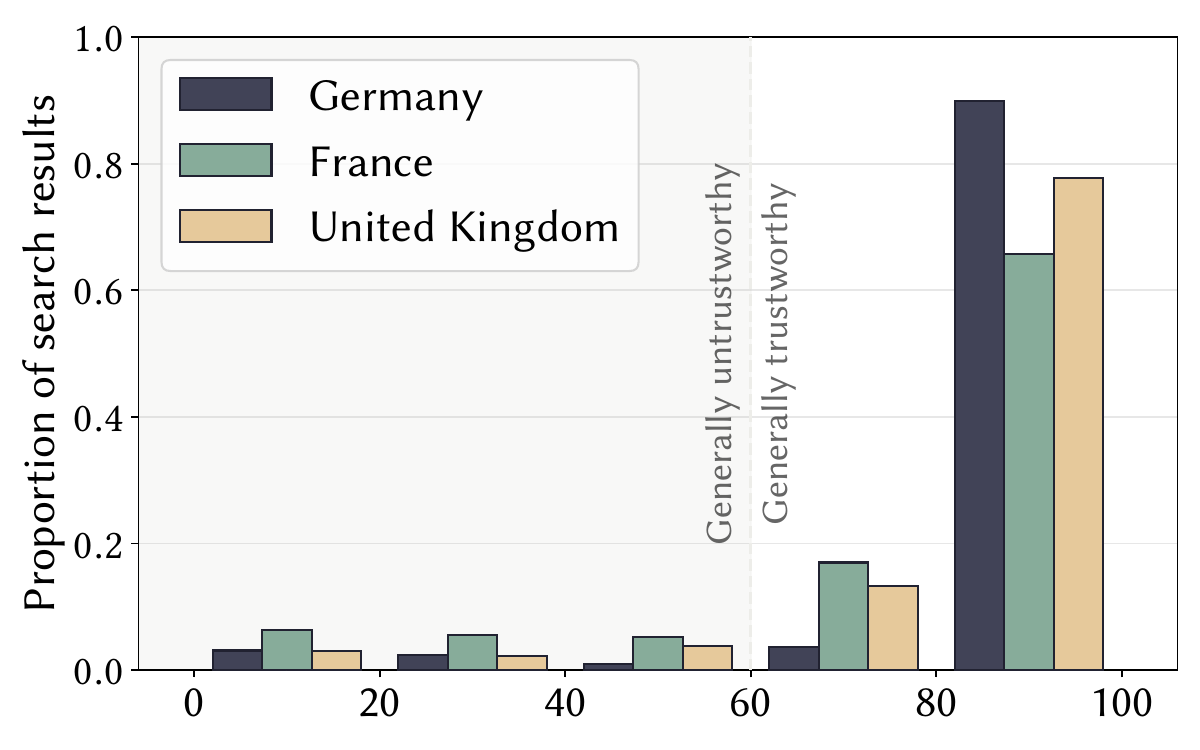}
        \caption{Credibility scores for queries about individuals.}
        \label{fig:credibility_distribution_per_country_people}
    \end{subfigure}
\caption{Search results per country returned by all four search engines combined. Panel (a) and (b) show how the political orientation of these search results are distributed per country, for queries about topics and individuals, respectively. A greater proportion of politicized content is seen in the UK in comparison to France and Germany, and there tends to be more left-leaning content for queries about individuals. Panel (c) and (d) show how the same search results are distributed according to credibility score, for queries about topics and individuals, respectively. These search results are in response to both neutral and negative queries. Political orientations and credibility scores are provided by NewsGuard (see Section \ref{sec:newsguard}).}
\label{fig:country_differences}
\end{figure}

Focusing first on the role of location, Figures \ref{fig:political_lean_per_country_general} and \ref{fig:political_lean_per_country_people} show how search results are distributed according to political orientation (characterized as either left-leaning, no bias, or right-leaning) in Germany, France, and the UK. These distributions, which are estimated from search results for all four search engines and for both neutral and negative queries, indicate that users in Germany, France, and the UK are exposed to slightly different proportions of politicized content. Those differences vary depending on whether the queries concern topics (Figure \ref{fig:political_lean_per_country_general}) or individuals (Figure \ref{fig:political_lean_per_country_people}). For queries about topics, we note that in France we found more right- than left-leaning content. In all three countries, however, the majority of the domains referenced by the search results lacked a political orientation. For queries about individuals, there is comparatively more left-leaning content. But again, the majority of search results were politically neutral. In the UK, though, this politically neutral majority amounts to only slightly more than half of the search results. 
Turning to the issue of exposure to low-credibility content, Figures \ref{fig:credibility_distribution_per_country_general} and \ref{fig:credibility_distribution_per_country_people} show how search results are distributed according to credibility scores (ranging from 0 to 100). Whether the query concerned a topic or an individual has little effect on the credibility distributions. Here, we find that the vast majority of search results refer to web domains with a credibility score greater than 60, which is considered the credibility threshold by NewsGuard. What this course-grained view on our results indicates is that the majority of search results delivered to users in the UK, France, and Germany are credible and, more often than not, hosted by domains that are not associated with politically partisan content. 

\subsection{Differences between search engines}\label{sediffs}

\begin{figure}[t]
    \centering
    \begin{subfigure}[t]{0.49\textwidth}
        \centering
        \includegraphics[height=3.8cm]{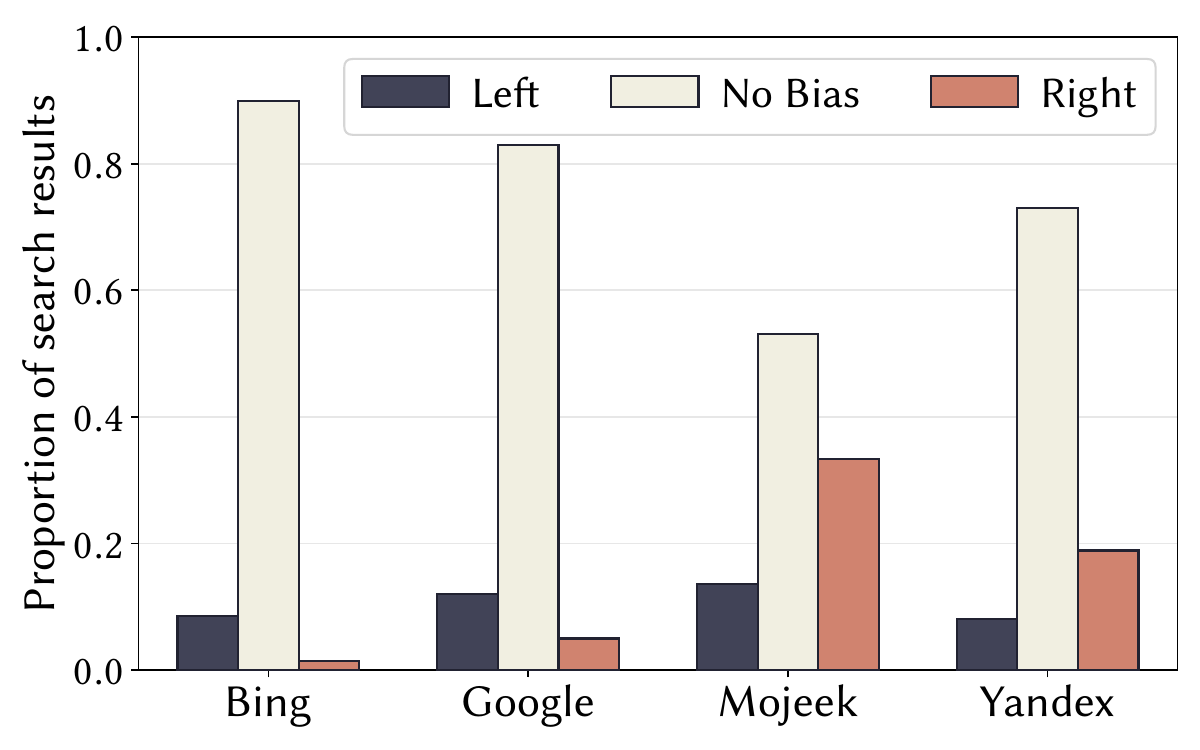}
        \caption{Political orientation for topics queries.}
        \label{fig:political_lean_per_se_general}
    \end{subfigure}%
    \begin{subfigure}[t]{0.49\textwidth}
        \centering
        \includegraphics[height=3.8cm, trim=32 0 0 0, clip]{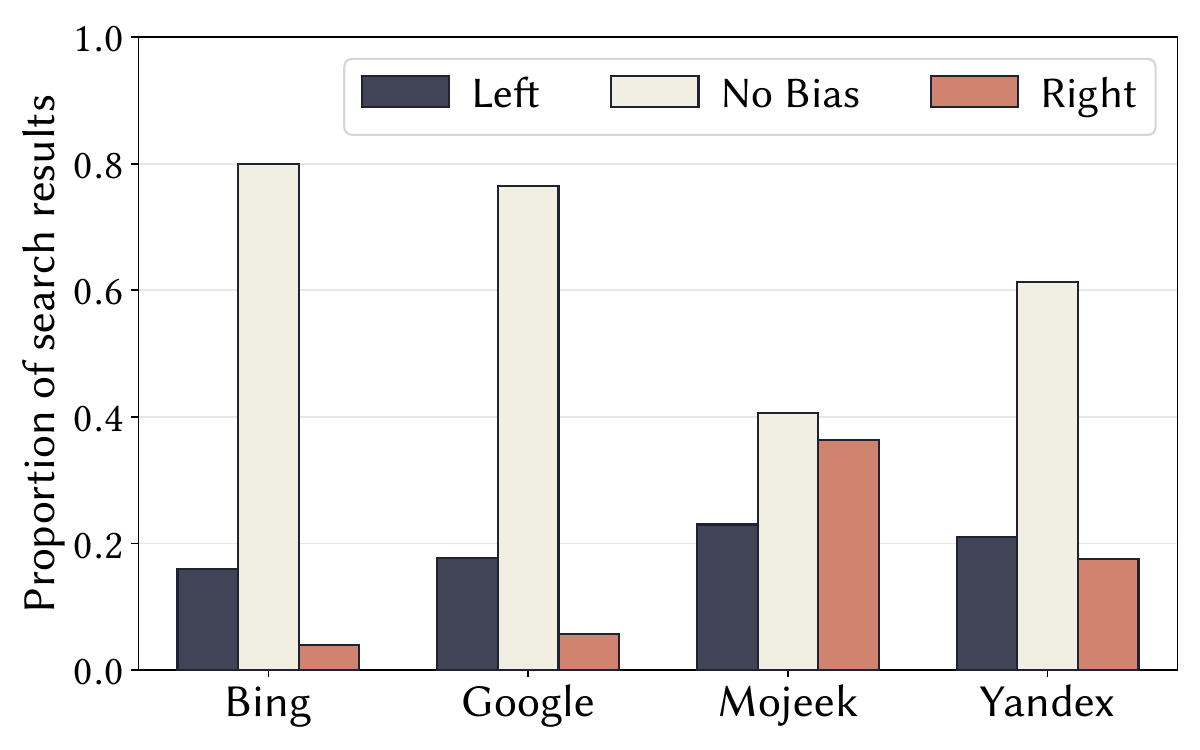}
        \caption{Political orientation for queries about individuals.}
        \label{fig:political_lean_per_se_people}
        \hspace{0.5cm}
    \end{subfigure}
    \begin{subfigure}[t]{0.49\textwidth}
        \centering
        \includegraphics[height=3.8cm]{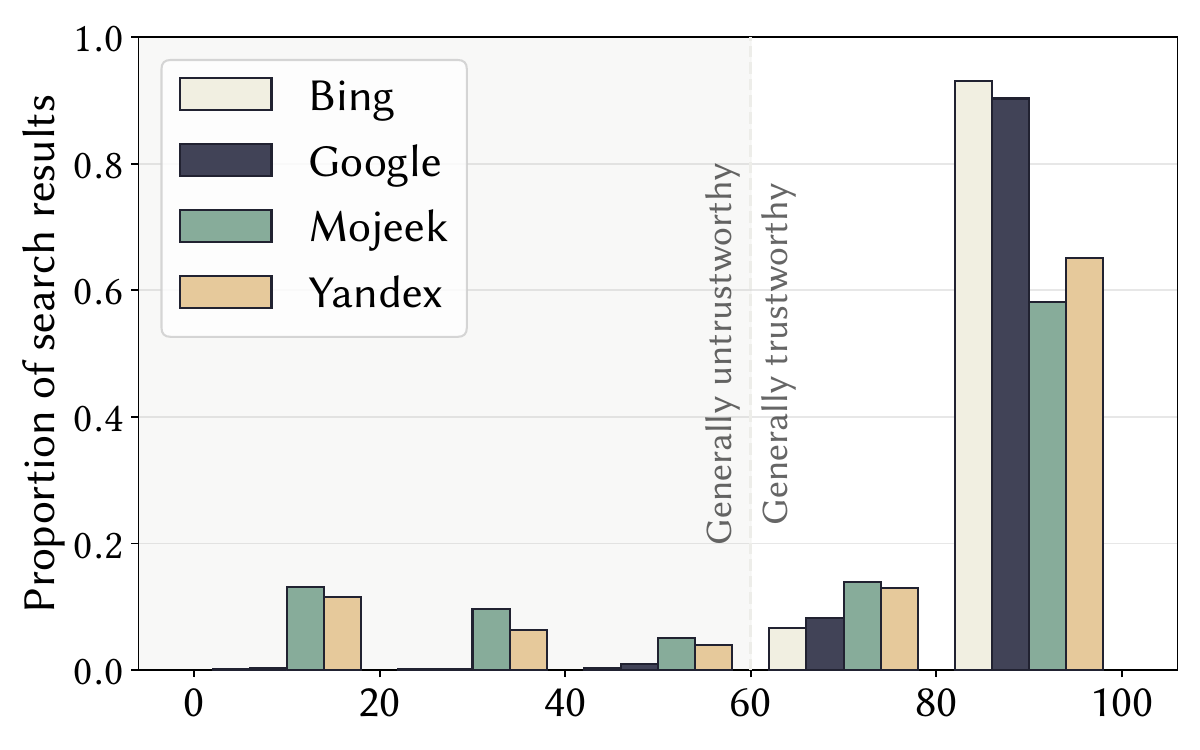}
        \caption{Credibility scores for topics queries.}
        \label{fig:credibility_distribution_per_se_general}
    \end{subfigure}
    \begin{subfigure}[t]{0.49\textwidth}
        \centering
        \includegraphics[height=3.8cm, trim=32 0 0 0, clip]{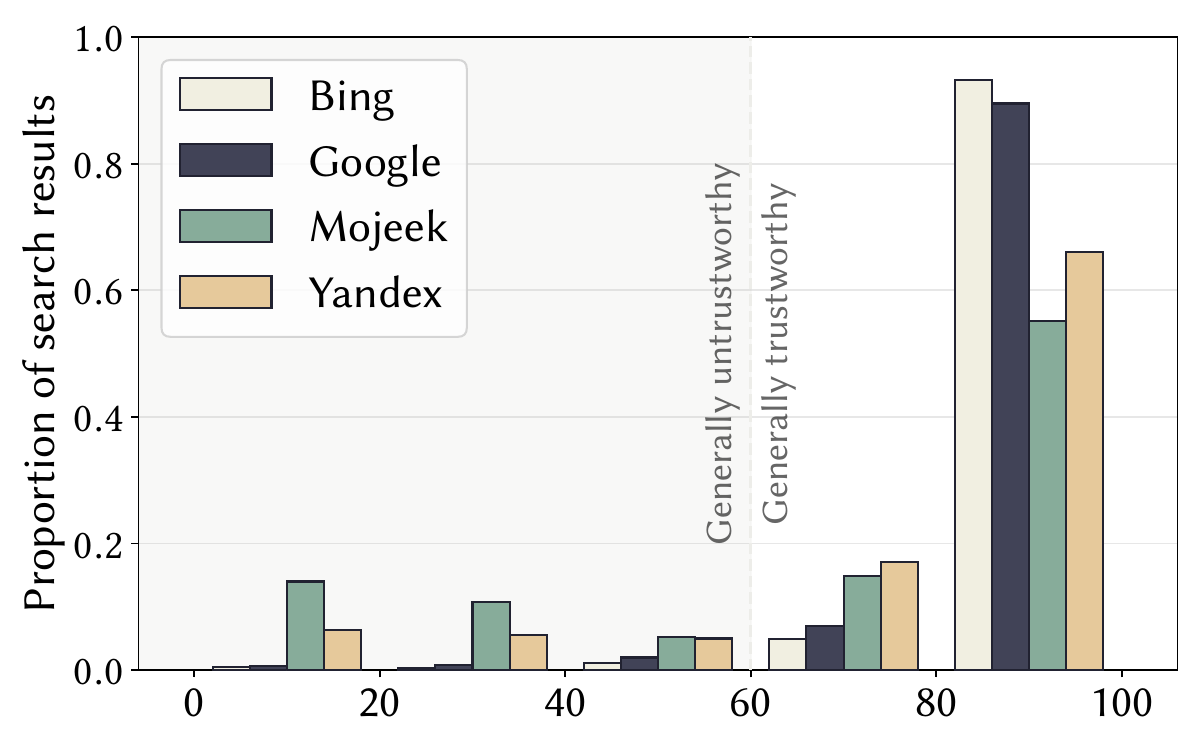}
        \caption{Credibility scores for queries about individuals.}
        \label{fig:credibility_distribution_per_se_people}
    \end{subfigure}
\caption{Focusing on search results for all three countries combined, panels (a) and (b) show how the political orientation of these search results are distributed by search engine, for queries about topics and queries about individuals, respectively. Panels (c) and (d) show how the same search results are distributed by credibility score. Mojeek and Yandex show more content from low credibility sources. These search results include responses to both neutral and negative queries.}
\label{fig:se_differences}
\end{figure}

In Figure \ref{fig:country_differences}, we analyzed the search results of all four search engines combined. In Figure \ref{fig:se_differences} we pay close attention to differences between the search engines, focusing again on the political orientation and credibility of the search results. 

\subsubsection{Political orientation} Figures \ref{fig:political_lean_per_se_general} and \ref{fig:political_lean_per_se_people} show that the search results returned by Bing had the highest proportion of sources with no political orientation (between 80\% and 90\%), followed by Google (between 77\% and 83\%). Both of these search engines return more left-leaning than right-leaning sources, and queries about individuals appear to amplify this tendency. In contrast, Mojeek and Yandex returned more results from politically slanted sources, surpassing Bing and Google in terms of the proportion of left-leaning content. Crucially, though, both Mojeek and Yandex show considerably more right-leaning content than left-leaning content (with the exception being Yandex, when queried about individuals). For both Mojeek and Yandex, nearly twice as many references to right-leaning outlets than left-leaning outlets are returned for queries about topics. What these results tell us is that the choice of search engine has a major impact on the degree of exposure to politicized content. Specifically, Google and Bing are behaviorally similar and tend to be biased to the left, whereas Mojeek and Yandex tend to return more politicized content than both Google and Bing, and tend to be biased to the right. 

\subsubsection{Credibility}\label{sec:exp1_three_types_low_cred_per_SE}
 Again, for all three countries combined, Figures \ref{fig:credibility_distribution_per_se_general} and \ref{fig:credibility_distribution_per_se_people} show how the same search results are distributed by credibility scores ranging from 0 to 100. Here we see only minor differences between queries about topics and queries about individuals, with the majority search results referencing web domains with credibility scores between 80-100. Notably, Bing and Google very rarely reference content with a credibility score below NewsGuard's trustworthiness threshold of 60. On the first page of search results, Bing returns a mean of 0.1 results from sources under this threshold, and Google a mean of 0.3 search results. In contrast, on the first page of search results, Mojeek references a mean of 2.9 search results and Yandex references a mean of 1.9 search results from sources below the credibility threshold. To better understand the prevalence of low-credibility and other suspicious content, we used NewsGuard's metadata to extract the proportions of search results that reference web content that either: (1) had a credibility score less than 60; (2) was hosted on domains known to promote conspiracy theories; or (3), was hosted on domains that have been flagged for publishing content relating to one or more forms of disinformation (see Table \ref{tab:newsguard_features}). 

\begin{figure}[t]
\centering
    \captionbox{For each search engine, the proportions of search results associated with one of 3 markers of problematic content: a credibility score below 60, a record of hosting conspiracy-related content, and a record of disinformation-related content. All markers are based on NewsGuard metadata. These results combine results from Germany, France, and the UK, and include search results relating to both neutral and negative topic queries. 
      \label{fig:proportions_problematic_content}}%
      [.47\linewidth]{%
      \includegraphics[height=4.0cm, trim=6 6 6 6, clip]{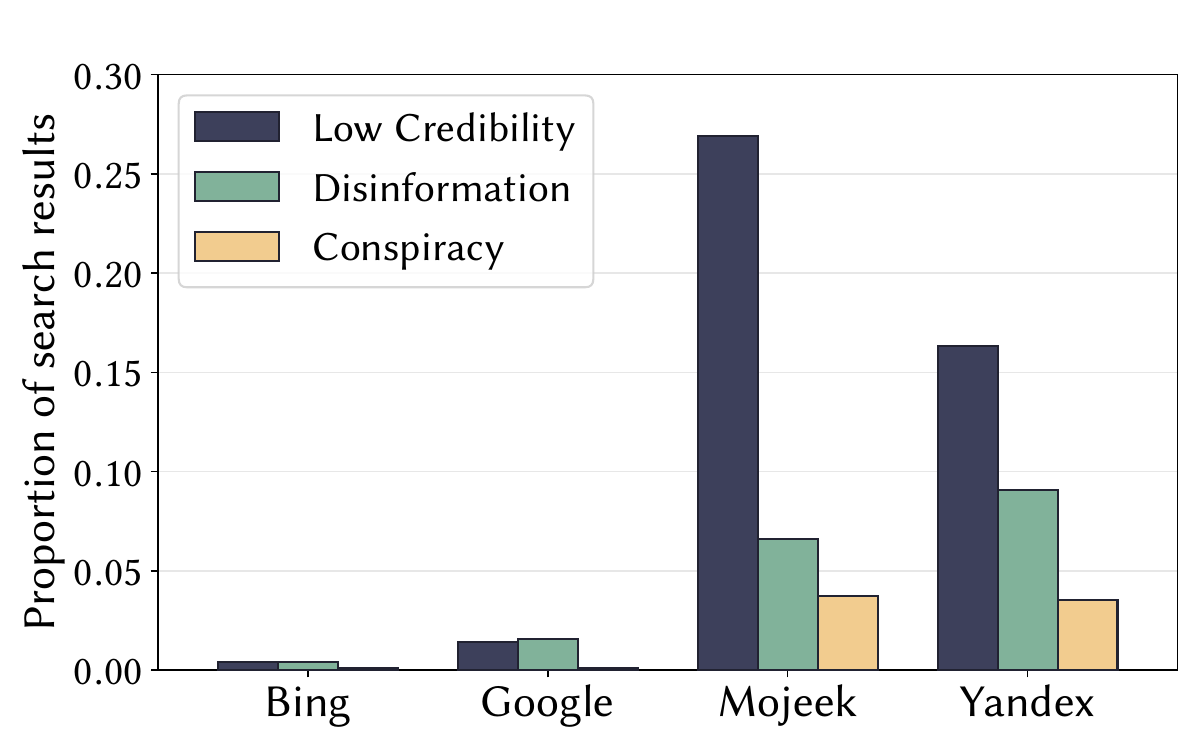}
      }
      \hfill
    \captionbox{For each search engine, the distribution of credibility scores for each political orientation label. These distributions have been normalized to account for differences in the absolute numbers for each orientation label and search engine. Appendix \ref{app:counts_per_se_label_and_cred_range} details these absolute numbers. 
      \label{fig:distrib_cred_over_pol}}
      [.47\linewidth]{%
      \includegraphics[height=4.15cm, trim=45 15 15 15, clip]{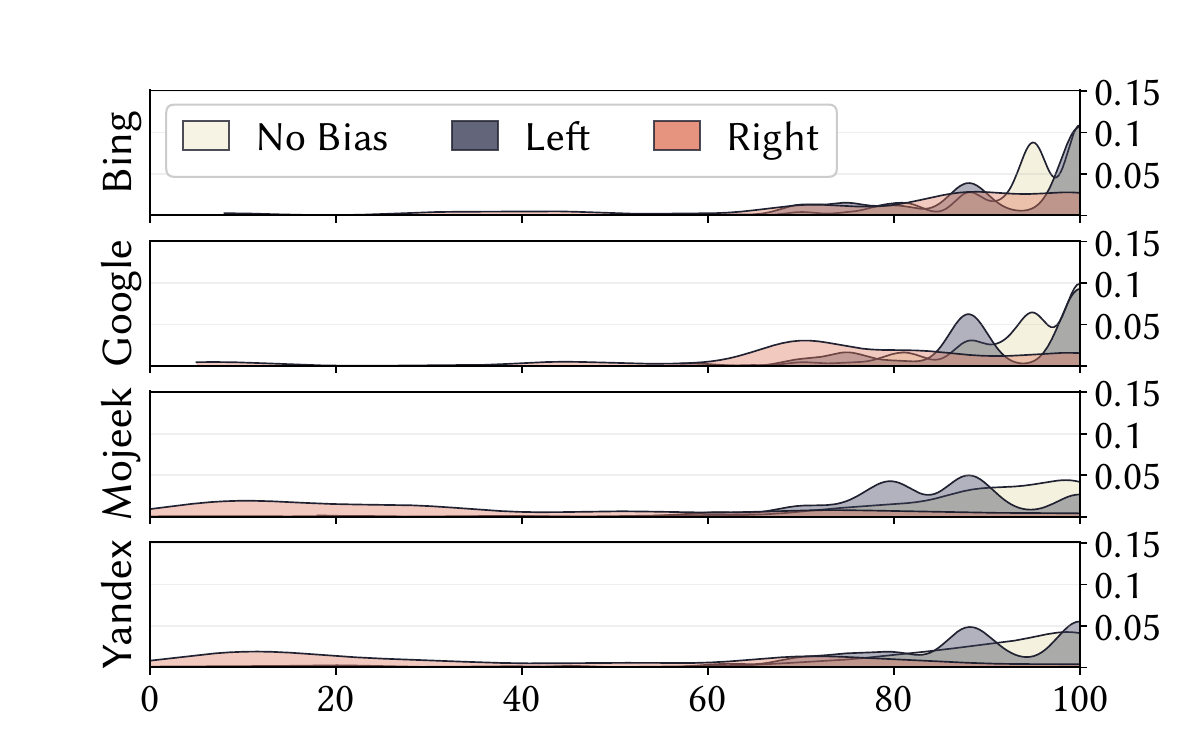}
      }
\end{figure}

For each of these three markers of low-credibility content, Figure \ref{fig:proportions_problematic_content} shows the proportion of search results returned by each search engine. Because these results are very similar for both queries about topics and queries about individuals, we will focus only on results relating to queries about topics. First, we see that less than 0.1\% of the results returned by Bing and Google reference web domains associated with conspiracy theories. For Google, only 1.6\% of results reference domains associated with disinformation, and for Bing this figure drops to less than 0.1\% of results. In contrast, approximately 3.8\% of the results returned by Mojeek and 3.6\% of the results returned by Yandex's reference domains associated with conspiracy theories. Focusing on domains associated with disinformation, 9\% of the results returned by Yandex have been flagged for disinformation, and for Mojeek this proportion is slightly lower at 7\%.

\subsubsection{Credibility scores and political orientation}
So far, we have viewed the credibility of a web domain and its political orientation as being unrelated. Is it the case, though, that low-credibility content is distributed equally across left leaning, right leaning, and apolitical web domains? For the search results returned by each search engine, Figure \ref{fig:distrib_cred_over_pol} shows the distribution of credibility scores for each of these respective political orientations. 
Because the results for queries about individuals and results for queries about topics are very similar, we will focus only on the results for queries about topics. 
First, notice how the credibility scores associated with right-leaning search results are widely distributed, and feature outlets with both very low and very high credibility scores. In contrast, left-leaning and politically neutral search results are concentrated above the credibility threshold. For Bing and Google, low-credibility content is overwhelmingly from right-leaning domains. In general, though, the majority of right-leaning content is hosted by domains that are above the threshold of credibility (roughly 16\% of right-leaning results returned by Google and Bing are from domains with low credibility scores, compared to under 1\% of domains for left-leaning and politically neutral search results). Strikingly, Mojeek and Yandex have a strong tendency to display low-credibility right-leaning content, with 72\% of right-leaning content referenced by Mojeek being labeled as low-credibility, and 63\% of right-leaning content referenced by Yandex being low-credibility. In summary, on the first page of search results, Bing and Google respectively display a mean of 0.1 and 0.3 results that are both low credibility and have a political bias. For Mojeek and Yandex, these frequencies increase to a mean of 2.6 and 1.5 results per page, respectively.

\subsection{The impact of query polarity}\label{qpolarity}

Recall that the query dataset was structured such that, for each query category, there is an approximately equal number of neutral and negative queries. The issue we consider now is to what extent this property of the queries, which we refer to as their polarity, influences exposure to potentially polarizing web content.

\subsubsection{Query polarity and political orientation}\label{sec:lollipop_political}

\begin{figure}[t]
    \begin{minipage}[t]{0.60\textwidth}
        \centering
        \includegraphics[width=\textwidth, trim=15 10 13 10, clip]{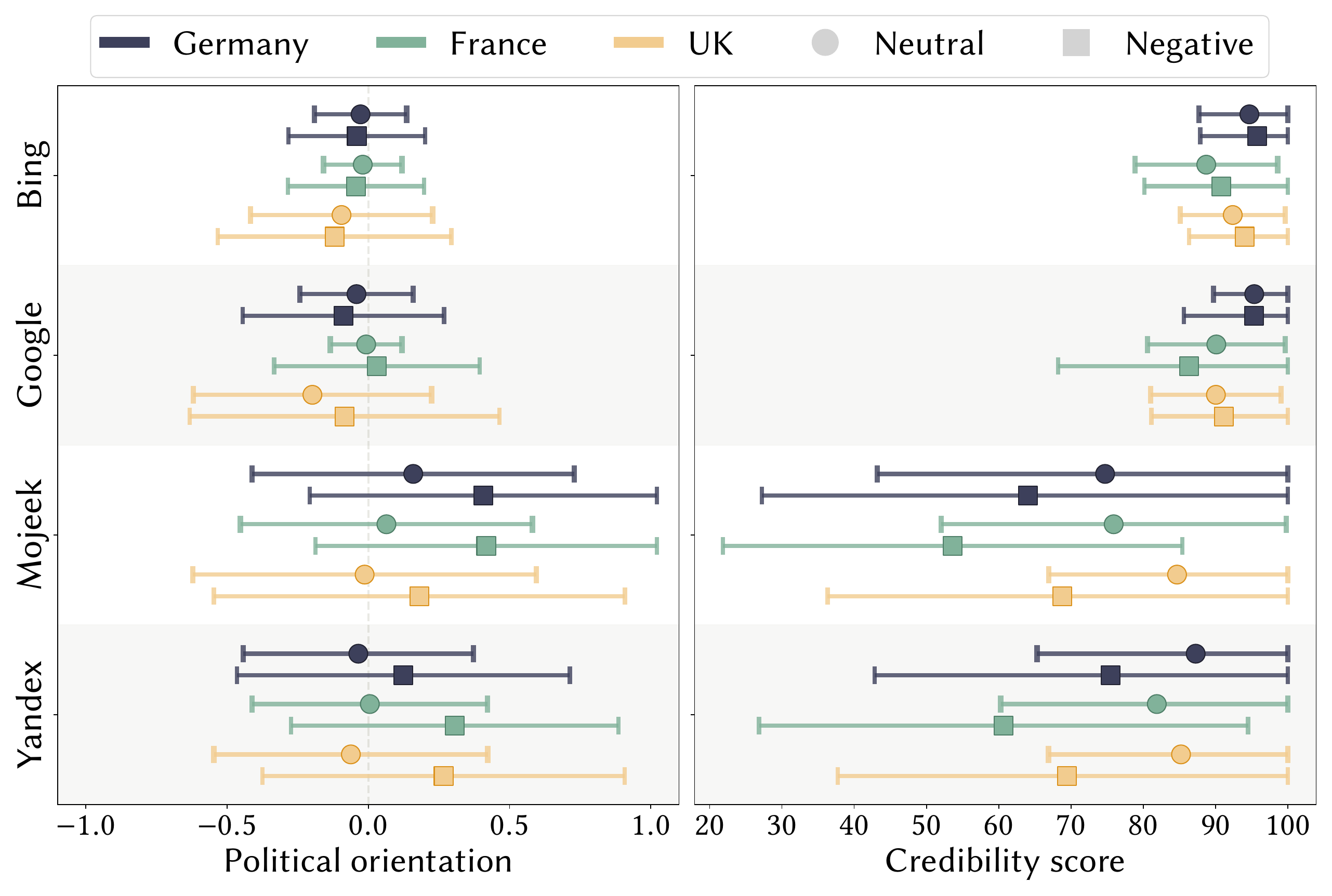}
        \caption{For each country, and for neutral and negative queries, the mean political orientation and credibility scores for queries about topics. The political orientation score assumes that left-leaning domains have a score of -1, right-leaning domains have a score of 1, and domains without a political orientation have a score of 0. Error bars represent standard deviations.}%
        \label{fig:lollipop_general}
    \end{minipage}%
    \hfill
    \begin{minipage}[t]{0.36\textwidth}
        \centering
        \includegraphics[width=0.792\textwidth, trim=10 -26 10 13, clip]{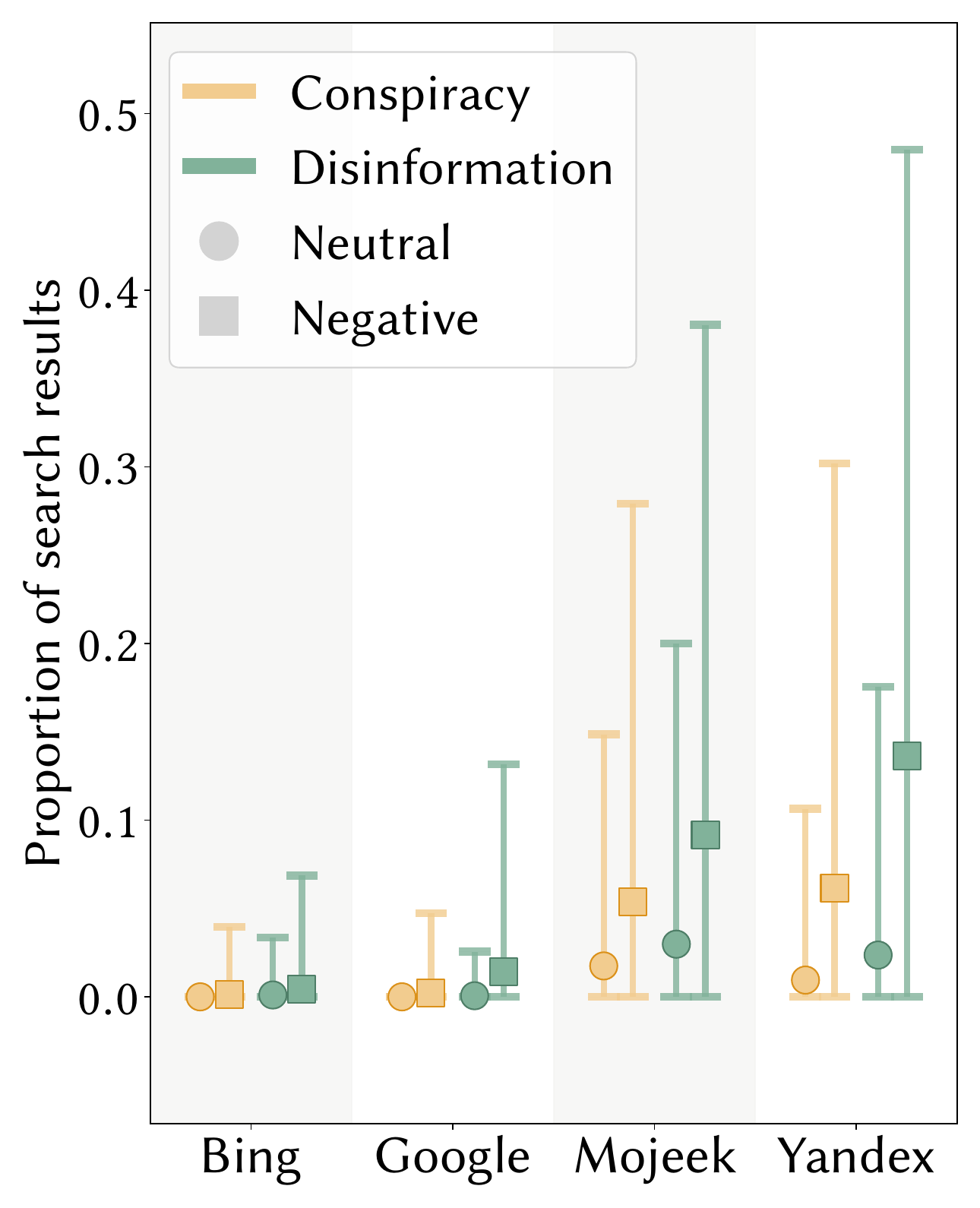}
        \caption{For each search engine, and for both query polarities relating to topics, the mean proportion of search results that reference conspiracy-related and disinformation-related web domains. Error bars represent standard deviations.}
        \label{fig:lollipop_disinfo_consp_general}
    \end{minipage}%
\end{figure}

\begin{figure}[t]
    \begin{minipage}[t]{0.60\textwidth}
        \centering
        \includegraphics[width=\textwidth, trim=15 10 13 10, clip]{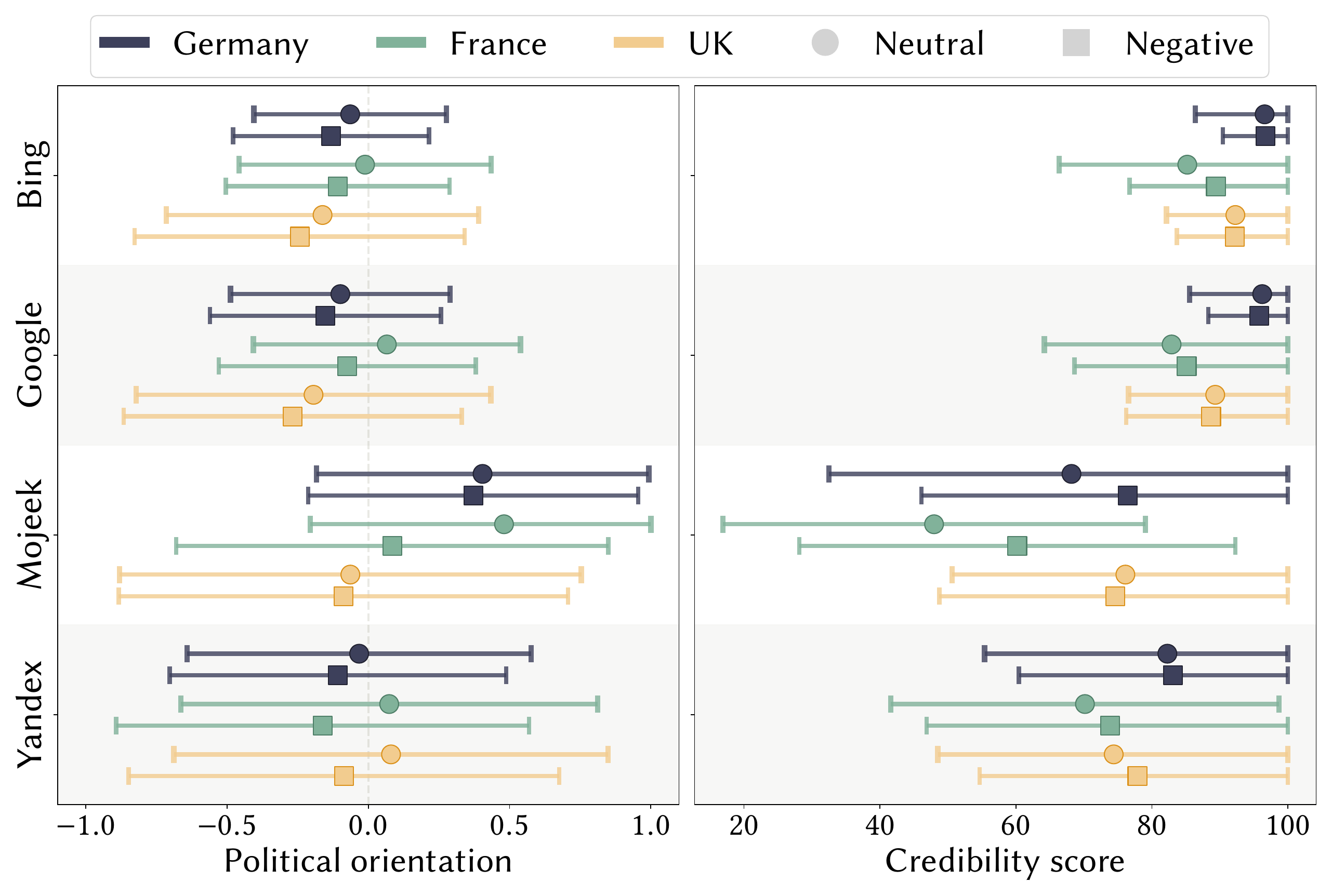}
        \caption{For each country, and for neutral and negative queries, the mean political orientation and credibility scores for queries about individuals. The political orientation score assumes that left-leaning domains have a score of -1, right-leaning domains have a score of 1, and domains without a political orientation have a score of 0. Error bars represent standard deviations.}%
        \label{fig:lollipop_people}
    \end{minipage}%
    \hfill
    \begin{minipage}[t]{0.36\textwidth}
        \centering
        \includegraphics[width=0.792\textwidth, trim=10 -26 10 13, clip]{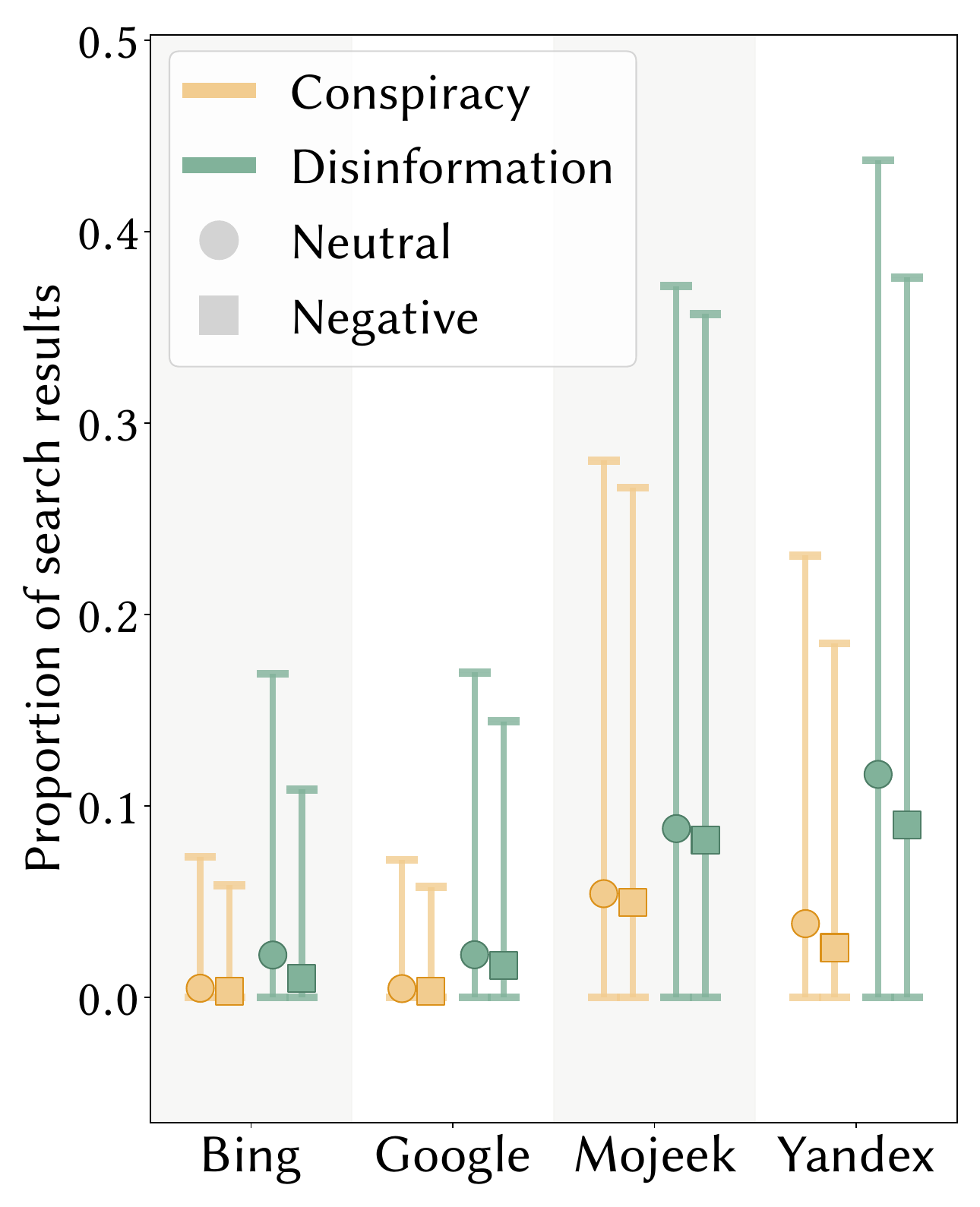}
        \caption{For each search engine, and for both query polarities relating to individuals, the mean proportion of search results that reference conspiracy-related and disinformation-related web domains. Error bars represent standard deviations.}
        \label{fig:lollipop_disinfo_consp_people}
    \end{minipage}%
\end{figure}

For each search engine in each of Germany, France, and the UK, Figure \ref{fig:lollipop_general} examines the mean political bias of the search results returned for both negative and neutral queries about topics (as opposed to individuals). We see that Google and Bing return results which are largely politically balanced, although with slight left-leaning bias for neutral queries, particularly in the UK. For both of these search engines, the polarity of the query had a relatively minor effect, although Google shows a slight shift towards right-leaning results for negative queries. Turning to Yandex and Mojeek, the variation in the political orientation of the search results is notably greater. Both search engines have a slight right-leaning bias that is particularly pronounced for negative queries. In short, the query polarity appears to have a consistently observed impact on the political orientation of the search results for Mojeek and Yandex, but less so for Bing and Google. Turning to queries about individuals, Figure \ref{fig:lollipop_people} shows, as we would expect, that for queries referencing individuals, query polarity has the opposite effect. All search engines shift towards left-leaning content when the queries are negative, because the negative polarity refers to a negative perspective on the individual, rather than LGBTIQ+ people.

\subsubsection{Query polarity and credibility}\label{sec:lollipop_credibility}
Turning to the relationship between query polarity and credibility, Figures \ref{fig:lollipop_general} and \ref{fig:lollipop_people} show that for queries about both topics and individuals, the query polarity has a relatively minor impact on the behavior of Bing and Google. However, topic queries result in a clear reduction in the mean credibility of the search results returned by Mojeek and Yandex. For queries about individuals, though, the query polarity either has little impact or slightly improves the mean credibility of the results. These differences, however, are minor in comparison to the differences that arise due to the choice of search engine, where there is a clear behavioral distinction between Google and Bing on the one hand, and Mojeek and Yandex on the other. Google and Bing consistently return search results referencing high-credibility domains, while Mojeek and Yandex return results that span high- and low-credibility domains.

\subsubsection{Query polarity and exposure to domains associated with conspiracy theories and disinformation}
Figure \ref{fig:lollipop_general} suggests that Mojeek and Yandex are more reactive to query polarity than Google and Bing, because they return results that differ in both political bias and credibility depending on the query polarity. Much like the analysis of Section \ref{sec:exp1_three_types_low_cred_per_SE}, in Figures \ref{fig:lollipop_disinfo_consp_general} and \ref{fig:lollipop_disinfo_consp_people} we examine the characteristics of low-credibility content arising from negative versus neutral queries in more detail. Specifically, we examine the proportions of search results from domains associated with content flagged by NewsGuard as containing disinformation or relating to conspiracy theories. For Mojeek and Yandex, Figure \ref{fig:lollipop_disinfo_consp_general} shows that search results returned for negative queries about topics contain more references to conspiracy-related web domains and domains associated with disinformation (particularly for Yandex). 
In numbers, negative queries issued to Mojeek increased the proportion of search results referencing domains associated with disinformation from 3\% to 9\%, while for Yandex the increase was from 2\% to 14\%. The number of conspiracy-related domains increase from 2\% to 5\% for Mojeek, and from 1\% to 6\% for Yandex. A very different picture emerges for Bing and Google. Bing shows less than 0.5\% of such content regardless of query polarity. Google shows slightly more disinformation for negative queries (1.4\% compared to 0.1\% for neutral queries). Once again, we see an opposite effect for queries about individuals in Figure \ref{fig:lollipop_disinfo_consp_people}. Neutral queries have higher amounts of conspiracy and disinformation content for all search engines (around as much as negative queries about LGBTIQ+ topics), and negative queries result in a drop of such problematic content, although the differences are minor.

\subsubsection{Query polarity and the ranking of search results}

\begin{figure}[ht]
    \centering
    \begin{subfigure}{1\textwidth}
    \includegraphics[width=\textwidth, trim=6 6 6 6, clip]{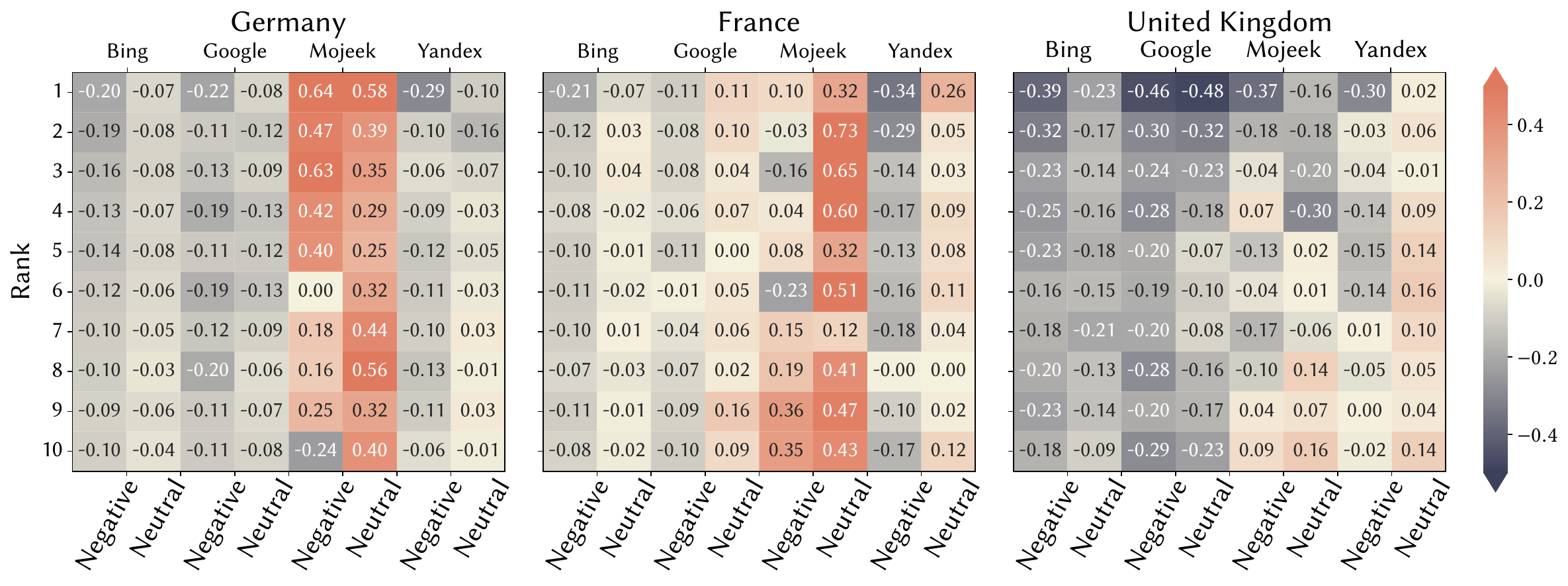}
    \caption{Political bias scores}
    \end{subfigure}
    \begin{subfigure}{1\textwidth}
    \vspace{0.2cm}
    \includegraphics[width=\textwidth, trim=6 6 6 6, clip]{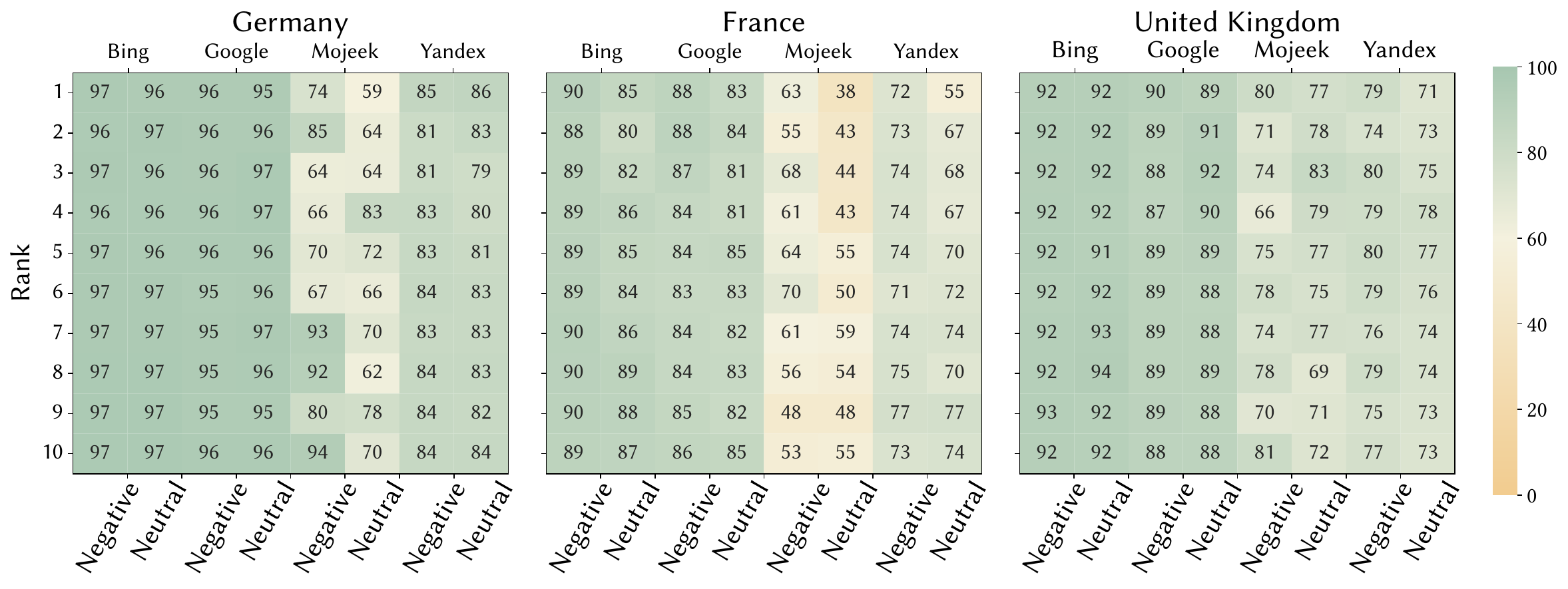}
    \caption{Credibility scores}
    \end{subfigure}%
    \caption{Examining the role of search result rankings. In (a), for each search engine and for each of Germany, France, and the UK, we consider the mean political orientation per search result rank for queries about individuals. In (b), we conduct the same comparison for credibility scores. As before, the political orientation score assumes that left-leaning domains have a score of -1, right-leaning domains have a score of 1, and domains without a political orientation have a score of 0.}
    \label{fig:rank_SERPS_people}
\end{figure}

The ranking of search results is a key consideration that we have so far glossed over. For instance, if politically biased or low-credibility search results tend to appear toward the bottom of the first page of results, then this is less concerning than if they tend to appear at the top. Indeed, previous research has identified cases where the ranking of search results is consistently biased toward one political orientation at the expense of another \cite{robertson_auditing_bias_2018}. To examine this possibility, we considered the first ten positions on the first page of results for each search engine. We then compared the mean political orientation and the mean credibility score of the results appearing at each of these positions. 
For queries about LGBTIQ+ topics, we found no systematic and consistent bias in the ranking of search results, for neither political bias scores nor credibility scores, and this absence of a consistent bias was found for both negative and neutral queries.\footnote{These findings are in fact consistent with previous work that found political bias in the ranking of search results, because these biases were only evident when looking beyond the first ten search results \cite{robertson_auditing_bias_2018}. Here, we focus on the first ten search results and also find no rank-sensitive biases.} 

However, when examining the political orientation of results for queries about individuals, we observe what appears to be some consistent patterns. For each search engine, Figure \ref{fig:rank_SERPS_people}a details the mean political bias score at each of the first 10 search result positions. Figure \ref{fig:rank_SERPS_people}(b) shows the same information for credibility scores. What we observe are slight rank-sensitive biases for: (1) negative queries issued to Bing in all three countries; (2) search results returned by Google in the UK, for both neutral and negative queries; (3) search results returned by Mojeek in Germany; and (4), negative queries returned by Yandex in France. In summary, the impact of rank is generally inconsistent, but can affect the political leaning of results to some degree.

\subsection{The impact of query category} \label{sec:lollipop_basecond}

\begin{figure}[ht]
    \centering
    \includegraphics[width=0.99\textwidth]{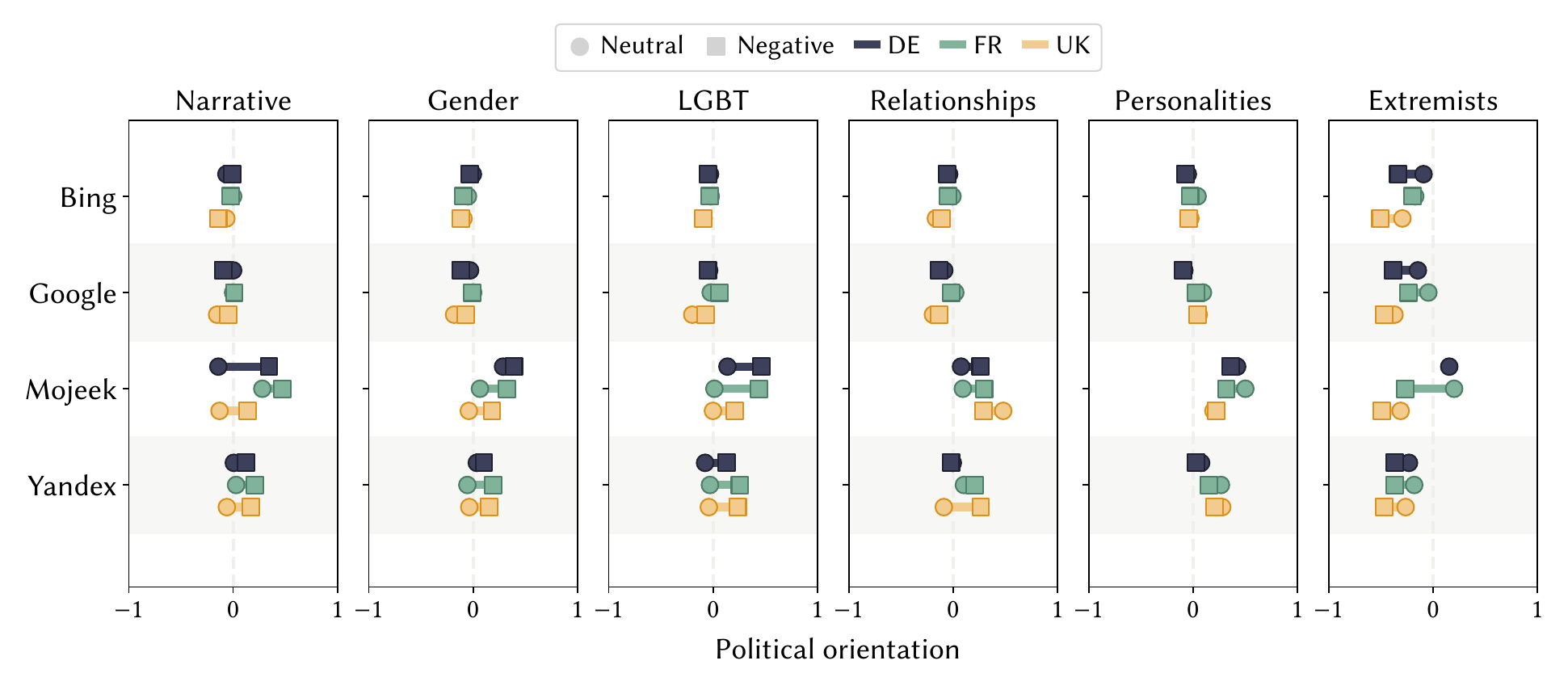}
    \caption{An analysis of all seven query categories. For each of the seven query categories, and for each search engine, we compare the mean political orientation of the search results in Germany, France, and the UK. The political orientation score assumes that left-leaning domains have a score of -1, right-leaning domains have a score of 1, and domains without a political orientation have a score of 0. See Appendix \ref{app:query_categories} for descriptions of the categories.}
    \label{fig:lollipop_basecond_bias}
\end{figure}

\begin{figure}[ht]
    \centering
    \includegraphics[width=0.99\textwidth]{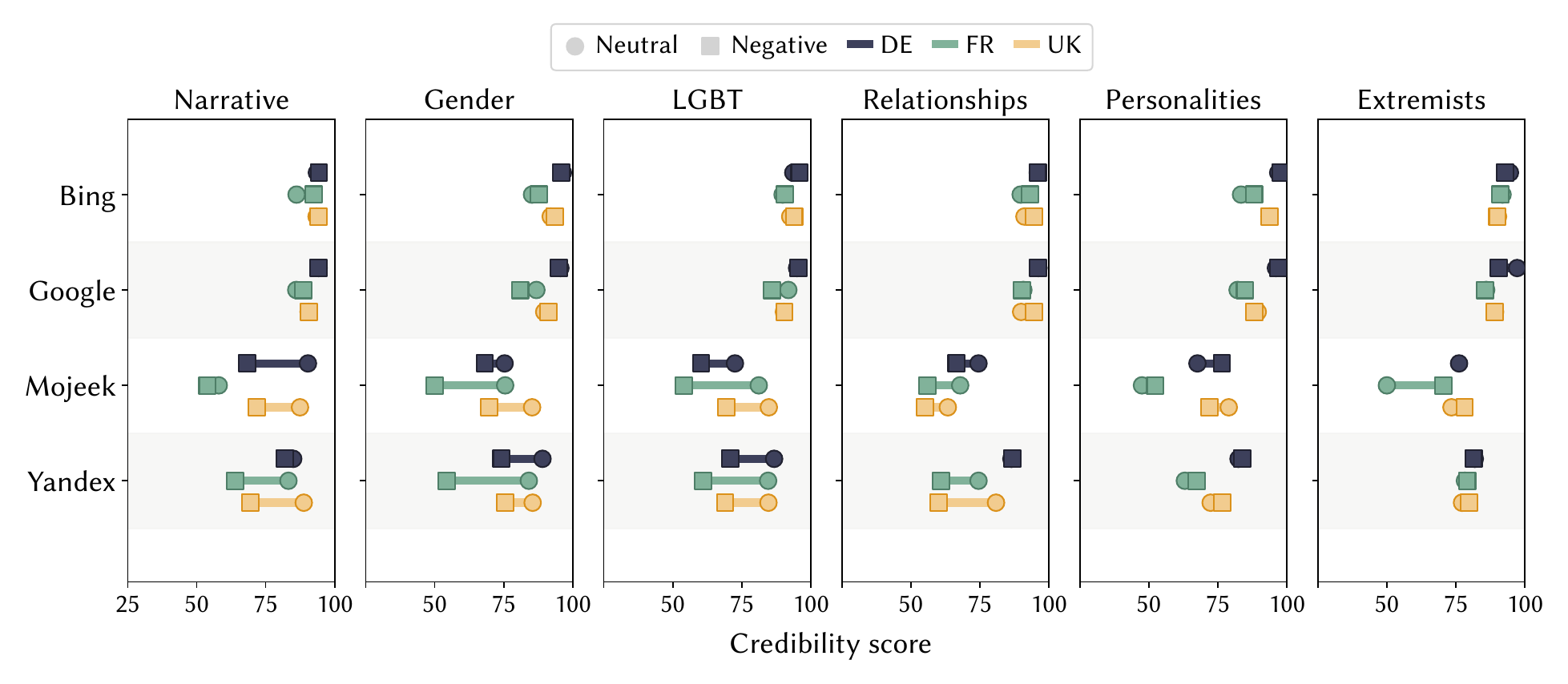}
    \caption{An analysis of all seven query categories. For each of the seven query categories, and for each search engine, we compare the mean credibility scores of the search results in Germany, France, and the UK. See Appendix \ref{app:query_categories} for descriptions of the categories.}
    \label{fig:lollipop_basecond_credibility}
\end{figure}

The final factor to consider is the impact of query category on exposure to polarizing content. Until now, our analysis of negative and neutral queries includes queries from all seven query categories detailed in Table \ref{tab:base_conditions}, which we have treated separately depending on whether the queries relate to LGBTIQ+ topics or to anti-LGBTIQ+ individuals. Focusing instead on all seven query categories individually, Figure \ref{fig:lollipop_basecond_bias} examines the mean political leaning of search results for each search engine, within each country. As before, we consider the impact of negative and neutral queries separately. For the first four categories (disinformation narratives, gender issues, LGBTIQ+ topics, and LGBTIQ+ relationships), we see a familiar pattern where Google and Bing are largely unreactive to the polarity of the query, while Mojeek and Yandex typically return more right-leaning and more low credibility content for negative queries. Figure \ref{fig:lollipop_basecond_credibility} conducts the same comparison for the final two query categories about individuals (anti-LGBTIQ+ personalities, and anti-LGBTIQ+ extremists). Here, we find that while both query categories show a left-leaning shift when queries are negative, the effect is larger and more consistent for anti-LGBTIQ+ extremists. On the other hand, the query polarity has little effect on the credibility of the results when considering queries about individuals.

\subsection{Queries that expose users to the greatest degree of polarizing content} \label{sec:polarisation_score}

In Section \ref{sec:se_overview} we characterized polarizing content as content that tends to stress ideological extremes, and often does so by deviating from mainstream journalist norms. 
Indeed, our analysis of polarizing content has considered both whether the content is politically partisan and whether it is of low credibility, as defined by NewsGuard. The final step in our analysis is to identify which queries expose users to the greatest degree of polarizing content by defining a polarization score that measures, for a particular search engine and a particular query, the degree to which the returned search results are of both low credibility and politically slanted. Specifically, for a given search engine, and for a given query $q$, we first calculate the mean credibility score of the returned search results, denoted $c_q$ (which ranges between 0 and 100). Because the query is likely to have been issued multiple times (e.g., from different locations), this mean is estimated from multiple pages of search results. From the same set of search results, we then calculate the mean political leaning, denoted $p_q$ (which ranges between -1 and 1, where -1 represents a bias to the political left and 1 represents a bias to the political right). The polarization score $P(q)$ of a query $q$ is then simply the product of the two quantities, after scaling the mean credibility score to range between 0 and 1: 
\begin{equation}
    P(q) = p_q \cdot (1 - \frac{c_q}{100}) %
    \label{eq:polarisation_score}
\end{equation}
The polarization score ranges from -1 (polarizing to the political left) to 1 (polarizing to the political right), and approaches zero when the search results tend to be of high credibility and are politically neutral. 

For each search engine, Table \ref{tab:most_polarising _queries} lists the three queries that yielded polarization scores with the greatest magnitudes, both to the political left (negative scores) and to the political right (positive values). First, notice that the queries with greatest absolute polarization scores (above 0.4) are polarizing to the political right, were issued to Mojeek and Yandex, and relate to topics rather than individuals. Google, on the other hand, yields polarization scores that are less than or equal to zero for all queries. Bing exhibits more varied behavior and yields low polarization scores in both political directions. To summarize, there are cases where all four search engines polarize to the left with low magnitude, but the greatest polarization occurs to the political right for Bing, Yandex, and Mojeek.

\begin{table}[t!]
\caption{For each search engine, the queries that resulted in the highest polarization scores, as defined by Equation \ref{eq:polarisation_score}. Negative scores indicate polarization to the political left, and positive scores indicate polarization to the political right.}
\label{tab:most_polarising _queries}
\footnotesize
\centering
    \begin{tabular}{lp{0.64\textwidth}r}
    \toprule
     &\textbf{Query}&\textbf{Polarization score}\\ [0.5ex]
    \midrule
    Bing & matt walsh is an extremist & -0.09\\
	 & bryan fischer is homophobic & -0.09 \\
	 & chaya raichik is homophobic & -0.06 \\
 	 & ... & ... \\
	 & ludovine de la rochère a raison & 0.04 \\
	 & ludovine de la rochère protège les valeurs & 0.08 \\
	 & ludovine de la rochère n'a pas peur d'appeler les choses par leur nom & 0.13 \\\hline
    Google & bryan fischer protects values & -0.07\\
	 & bryan fischer is an extremist & -0.05 \\
	 & transgender & -0.05 \\
 	 & ... & ... \\
	 & cyril hannouna est homophobe & 0.0 \\
	 & aymeric chauprade est extrémiste & 0.0 \\
	 & aymeric chauprade a raison & 0.0 \\\hline
    Mojeek & chaya raichik is homophobic & -0.09\\
	 & trans people were not persecuted by the nazi regime & -0.03 \\
	 & julia hartley brewer is an extremist & -0.02 \\
 	 & ... & ... \\
	 & transgender & 0.13 \\
	 & aymeric chauprade a raison & 0.44 \\
	 & surgically mutilate trans confused kids & 0.65 \\ \hline
    Yandex & roy moore is an extremist & -0.06\\
	 & roy moore ist transphobisch & -0.06 \\
	 & andrzej duda ist lgbt feindlich & -0.01 \\
 	 & ... & ... \\
	 & trans prison rape & 0.12 \\
	 & andrew tate ist transphobisch & 0.14 \\
	 & lgbt lobby & 0.42 \\
\bottomrule
\end{tabular}
\end{table}

\section{Discussion} \label{sec:discussion}

To what extent do search engines expose users to polarizing content? To examine this question, we monitored four independent search engines in eight European countries over a three-month period, and gathered over 1.5 million search results. Using the NewsGuard dataset, our analysis has focused on search engine behavior in the UK, France, and Germany. We have characterized polarizing content as content that, according to NewsGuard, has both low-credibility and is politically slanted. For queries relating to LGBTIQ+ people, we have examined how location, query category, query polarity, and choice of search engine contribute to exposure to potentially polarizing content. Our results show that, although all search engines have the capacity to return results that reference polarizing content, the frequency with which users are likely to encounter this content is determined in large part by their choice of search engine. The location of the user and the composition of the query play a relatively minor role. Before considering the implications and limitations of our findings, we will first outline what they mean, in simple quantitative terms, for search engine users. 

\subsection{Factors influencing people's exposure to polarizing content}

In numbers, our results (detailed in Section \ref{sediffs}) indicate that a Google user who switches to Mojeek can expect to encounter over 8 times as many search results that reference content that is both low-credibility and politically partisan. A Bing user switching to Mojeek can expect to see approximately the same increase. In absolute terms, Google returns a mean of less than 1 potentially polarizing search result on the first page of results (3 in every 100 search results), while Mojeek returns a mean of over 2 potentially polarizing results on the first page of results (26 in every 100 search results). When switching to Yandex, Google and Bing users can expect to encounter at least 5 times as many results that reference potentially polarizing content. In short, we have observed a clear behavioral distinction between Bing and Google on the one hand, and Mojeek and Yandex on the other. This quantitative summary relates only to searches about LGBTQI+ people and topics, and is based on results that have been aggregated over languages, locations, and query polarities. Could it be that these aggregated findings mask additional factors that search engine users should be aware of? While other factors certainly play a role, they are of comparatively minor importance. For example, if a user of Google or Bing switches from issuing only neutral queries to issuing only negative queries in relation to LGBTQI+ people, this shift would result in no discernible difference in their exposure to polarizing content. For Mojeek and Yandex users, though, this same behavioral shift would lead to a consistent but minor difference. Specifically, they should expect see an increase from approximately 13 polarizing results in every 100 first page search results to approximately 16 in 100 (see Section \ref{qpolarity}). The impact of language and location, and differences resulting from the choice of query topic result in similarly minor differences. As we have shown, some of these differences appear to lack a systematic pattern, which suggests that in order to make firmer claims, a larger-scale study focusing in greater detail on factors such as query topic would be needed. Nevertheless, it is clear that the search engine a user chooses to use can influence their exposure to polarizing content to a far greater extent than what they search for, the language they use, or where they are located.

\subsection{Study limitations}

Rather than choose to passively capture real-world search engine interactions that involve human users with a history of online behavior, we have opted to automate and actively control all search engine interactions. Because these automated interactions are not linked to any user accounts, our approach cannot shed light on 
personalized search results. This might be seen as a disadvantage, because for search engines such as Google, real-world users could have been exposed to different search results to those that we have collected and analyzed. The advantage of our approach, though, is that we could systematically control the queries, locations, and most importantly, the choice of search engine. If we took a passive monitoring approach, where queries are spontaneously created by users during course of their everyday online activities, a controlled analysis of query polarity would be infeasible. Similarly, monitoring search engines with a minimal market share, like Mojeek, is unrealistic within a passive monitoring framework. Indeed, our focus on understanding the basic ability of search engines to deliver ranked results means that we can compare search engines on an equal footing: we have factored-out those design features that differ between search engines, such as whether they provide personalized results, and how they adapt the page layout to include, for example, separate sections detailing related news stories and social media posts.

Our analysis rests on some simplifying assumptions. First, we have assumed that the (approximately) 12,000 domains documented by NewsGuard provide a representative and sufficiently large sample of web domains from which to draw our conclusions. Second, we used this dataset to infer document-level features (those relating to a document referenced by a search result) from domain-level features (those relating to a web domain that hosts potentially many documents). These assumptions are to a certain extent unavoidable due to a lack of reliable, web-scale datasets detailing the journalist practices and political biases associated with many hundreds of thousands of web resources \cite{ronnback2025automatic}. This is why previous search engine monitoring studies have made the same, or very similar, assumptions (e.g., \cite{robertson_auditing_personalization_2018,robertson_auditing_bias_2018,robertson_2023_users}). One implication of these assumptions is that our findings are likely to represent a best-case scenario. Specifically, NewsGuard has a tendency to document domains with significant audience exposure, and includes an approximately balanced distribution of politically left- and right-leaning web domains. This means that while the larger and well-known news outlets are well represented, the long tail of lesser known and more obscure web domains is under-represented. If we then assume that polarizing content is more likely to originate from domains occupying the long tail, then our analysis is likely to have under-estimated the prevalence of polarizing content in search results. Another factor that limits the scope of our conclusions is that we have only studied how search engines respond to queries about LGBTIQ+ topics and people. Extrapolating our conclusions to other ideologically sensitive social issues is an inference that would require additional, and targeted, monitoring studies. Similarly, we have analyzed the issue of polarization, which is a social phenomena that is far more complex and nuanced than our analysis of only two aspects, namely source credibility and political slant \cite{dimaggio_1996_have,weismueller_falsehood_partisanship_for_polarisation_2024}.

\subsection{Explaining search engine behavior}

In light of our findings, it is natural to ask why Google and Bing return very few low-credibility, politicized results in comparison to both Yandex and Mojeek. What causes these systematic differences? Because the algorithms used by search engine technologies are hidden from public view, any attempt to answer these questions will necessarily be speculative. Such speculation can nevertheless make use of information in the public domain. Mojeek, for example, is a relatively small UK operation with the expressed goal of differentiating itself from more established search engines by prioritizing web content from less well-known and less authoritative sources \cite{mojeekIndependentResults}. Yandex, on the other hand, is a much larger and well-established operation based in Russia, and has repeatedly been accused of Kremlin influence \cite{kravets2024yandex}, disregard for privacy concerns \cite{gerbet_privacy_analysis_Google_Yandex}, censorship \cite{Kravets_Yandex_news_bias}, and the promotion of misinformation \cite{kravets2024yandex}. We have found 
that these two search engines, Mojeek and Yandex, despite operating under very different circumstances, expose their users to similar degrees of potentially polarizing content. Given our findings, perhaps the least speculative explanation is that the differences between search engines we observe may reflect a limited range of opportunities for offering a substantive alternative to Google and Bing.  Google and Bing excel at delivering relevant results from high-credibility sources. To compete with Google and Bing, and offer something that these services lack, the path of least resistance may be to deliver relevant results, but from a greater number of low-credibility sources, and thereby cater to a growing audience that trusts and seeks information from sources outside of the mainstream media \cite{edelman2024,edelman2025}.

\section{Acknowledgements}
This research is part of the project ``The dynamics of LGBTQ+ misinformation across Europe'' which was funded by a grant from the European Media and Information Fund (\url{https://gulbenkian.pt/emifund/}). All inquiries should be directed to the principal investigator, Henry Brighton (\href{mailto:h.j.brighton@tilburguniversity.edu}{\nolinkurl{h.j.brighton@tilburguniversity.edu}}).

\bibliographystyle{acm}
\bibliography{bibliography}

\newpage
\appendix

\section{Appendix: Search Engine Localization Options}\label{app:localization}

\begin{table}[ht]
    \scriptsize
    \caption{Localization options used per search engine and country, as well as the four major cities where data was collected.}
    \label{tab:data_collection_appendix}
    \centering
    \begin{tabular}{m{0.065\linewidth} p{0.27\linewidth} p{0.27\linewidth} p{0.27\linewidth}}
        \toprule
         & \textbf{France} \newline(Bordeaux, Marseilles, Paris, Strasbourg )& \textbf{Germany} \newline (Berlin, Cologne, Dresden, Munich) & \textbf{United Kingdom} \newline(Belfast, Cardiff, Edinburgh, London)\\ 
         \midrule
         Bing &  \href{https://www.bing.com/?setlang=fr\&cc=fr\&cc=FR}{\tt .com/?setlang=fr\&cc=fr\&cc=FR} &  \href{https://www.bing.com/?setlang=de\&cc=de\&cc=DE}{\tt .com/?setlang=de\&cc=de\&cc=DE} & \href{https://www.bing.com/?setlang=en\&cc=GB}{\tt .com/?setlang=en\&cc=GB} \\
         Google  &  \href{https://www.google.fr/}{\tt .fr} &  \href{https://www.google.de/}{\tt .de} & \href{https://www.google.co.uk/}{\tt .co.uk}\\
         Mojeek & \href{https://www.mojeek.com/?reg=fr\&lb=fr\&arc=fr}{\tt .com/?reg=fr\&lb=fr\&arc=fr} &  \href{https://www.mojeek.com/?reg=de\&arc=de\&lb=de}{\tt .com/?reg=de\&arc=de\&lb=de} & \href{https://www.mojeek.com/?reg=uk\&lb=en\&arc=gb}{\tt .com/?reg=uk\&lb=en\&arc=gb}\\
         Yandex &  \href{https://yandex.com}{\tt .com}&  \href{https://yandex.com/}{\tt .com} & \href{https://yandex.com/}{\tt .com} \\ \bottomrule
    \end{tabular}
\end{table}

\section{Appendix: Query Categories}\label{app:query_categories}

Each query belongs to an overarching category. Table \ref{tab:base_conditions} provides a brief description of what each category entails. Additionally, there are varying numbers of unique queries per country, topic category, and query polarity (neutral or negative). The table \ref{tab:query_num_overview} provides an overview. 

\begin{table}[ht]
    \centering
    \footnotesize
    \caption{Overview of the categories of queries, with some representative examples.}
    \begin{tabular}{p{0.15\textwidth} p{0.40\textwidth} p{0.35\textwidth}} \toprule
         \textbf{Category}& \textbf{Description} &\textbf{Example}\\ \midrule
         Narratives & Query prompts search engine about common disinformation narratives. These are predominantly related to LGBT-related disinformation, but also include coronavirus- and vaccine-disinformation. &gays will traffic children, doctors and pharma forcing children into transition\\ \hline 
         Gender&  Query relates to traditional and other genders, homo- and transphobia.&gender dysphoria, 76 genders, transgender quotas\\ \hline 
         LGBT &  Query concerns topics surrounding LGBTIQ+ people, such as important events, discrimination and general information. &violence against trans people, homo propaganda, lgbt rights, lgbt meaning \\ \hline 
         Relationships&  Query relates to family relationships, particularly the relationship between parents and children.&same-sex parenthood, same sex couples can't bring up children correctly, surrogate motherhood and lgbt\\ \hline 
         Personalities &  Query is about a specific individual with some anti-LGBTIQ+ stances, particularly whether they are considered as being correct, or problematic.&jk rowling speaks the truth, allison pearson is homophobic, lord moonie protects values\\ \hline 
         Extremist &  Query relates to an individual that have been designated by the Southern Poverty Law Centre (SPLC) as extremists.&matt walsh speaks the truth, david lane is homophobic, fred phelps is a fascist\\\bottomrule
    \end{tabular}
    \label{tab:base_conditions}
\end{table}

\begin{table}[ht]
    \centering
    \caption{Overview of number of queries per topic category, country, and query polarity.}
    \footnotesize
    \begin{tabular}{ccccccc} \toprule 
         &  \multicolumn{2}{c}{\textbf{France}}&  \multicolumn{2}{c}{\textbf{Germany}}&  \multicolumn{2}{c }{\textbf{United Kingdom}}\\ 
    \textbf{Category} & Neutral& Negative& Neutral& Negative& Neutral&Negative\\
        \midrule
         Narratives &  7&  8&  13&  17&  9& 52\\ 
         Gender&  10&  10&  35&  31&  29& 53\\ 
         LGBT&  43&  34&  65&  71&  46& 71\\ 
         Relationships&  16&  14&  24&  22&  9& 25\\ 
         Personalities&  36&  36&  93&  93&  66& 66\\ %
         Extremist&  39&  39&  39&  39&  39& 39\\ 
         \bottomrule
    \end{tabular}
    \label{tab:query_num_overview}
\end{table}

\section{Appendix: Supplementary Data Exploration}

\subsection{NewsGuard Coverage}\label{app:ng_coverage}

Our data focuses on a relatively specific topic that may be under-represented in the NewsGuard data. To examine whether there is sufficient coverage by the NewsGuard metadata, we provide tables of the amounts of search results and unique web-domains with or without labels, both per country (see Table \ref{table:coverage_per_Country}) and per search engine (see Table \ref{table:coverage_per_SE}). 

\begin{table}[ht]
    \footnotesize
    \centering
    \caption{Coverage of NewsGuard per country}
    \label{table:coverage_per_Country}
     \begin{tabular}[b]{l R{0.12\linewidth}  R{0.12\linewidth}  R{0.12\linewidth}   R{0.12\linewidth}  R{0.13\linewidth}  R{0.12\linewidth}} \toprule
    & \textbf{Search results with metadata} & \textbf{Search results without metadata} & \textbf{Percentage of search results with metadata} & \textbf{Unique domains with metadata} & \textbf{Unique domains without metadata} & \textbf{Percentage of domains with metadata} \\
    \midrule
    DE & 82732 & 137352  & 37.6\%     & 692  & 6299    & 9.9\%      \\
    FR & 43608 & 77298   & 36.1\%     & 539  & 4422    & 10.9\%     \\
    UK & 88345 & 131982  & 40.1\%     & 949  & 7094    & 11.8\%  \\
    \bottomrule
    \end{tabular}
\end{table}

\begin{table}[ht]
    \caption{Coverage of NewsGuard per search engine.}
    \label{table:coverage_per_SE}
    \footnotesize
    \centering
    \begin{tabular}[b]{l R{0.11\linewidth}  R{0.11\linewidth}  R{0.12\linewidth}  R{0.11\linewidth}  R{0.12\linewidth}  R{0.12\linewidth}}
        \toprule 
        & \textbf{Search results with metadata} & \textbf{Search results without metadata} & \textbf{Percentage of search results with metadata} & \textbf{Unique domains with metadata} & \textbf{Unique domains without metadata} & \textbf{Percentage of domains with metadata} \\
        \midrule 
        Bing   & 77290 & 75698   & 50.5\%     & 761  & 3901   & 16.3\%     \\
        Google & 47333 & 88174   & 34.9\%     & 697  & 4714   & 12.9\%     \\
        Mojeek & 29440 & 107922  & 21.4\%     & 691  & 7107   & 8.9\%      \\
        Yandex & 60622 & 74838   & 44.8\%     & 1013 & 5153   & 16.4\%    \\
        \bottomrule
    \end{tabular}
\end{table}

\begin{figure}[ht]
\centering
\begin{minipage}[t]{0.60\linewidth}
  \centering
  \includegraphics[height=5.3cm, trim=10 5 5 5, clip]{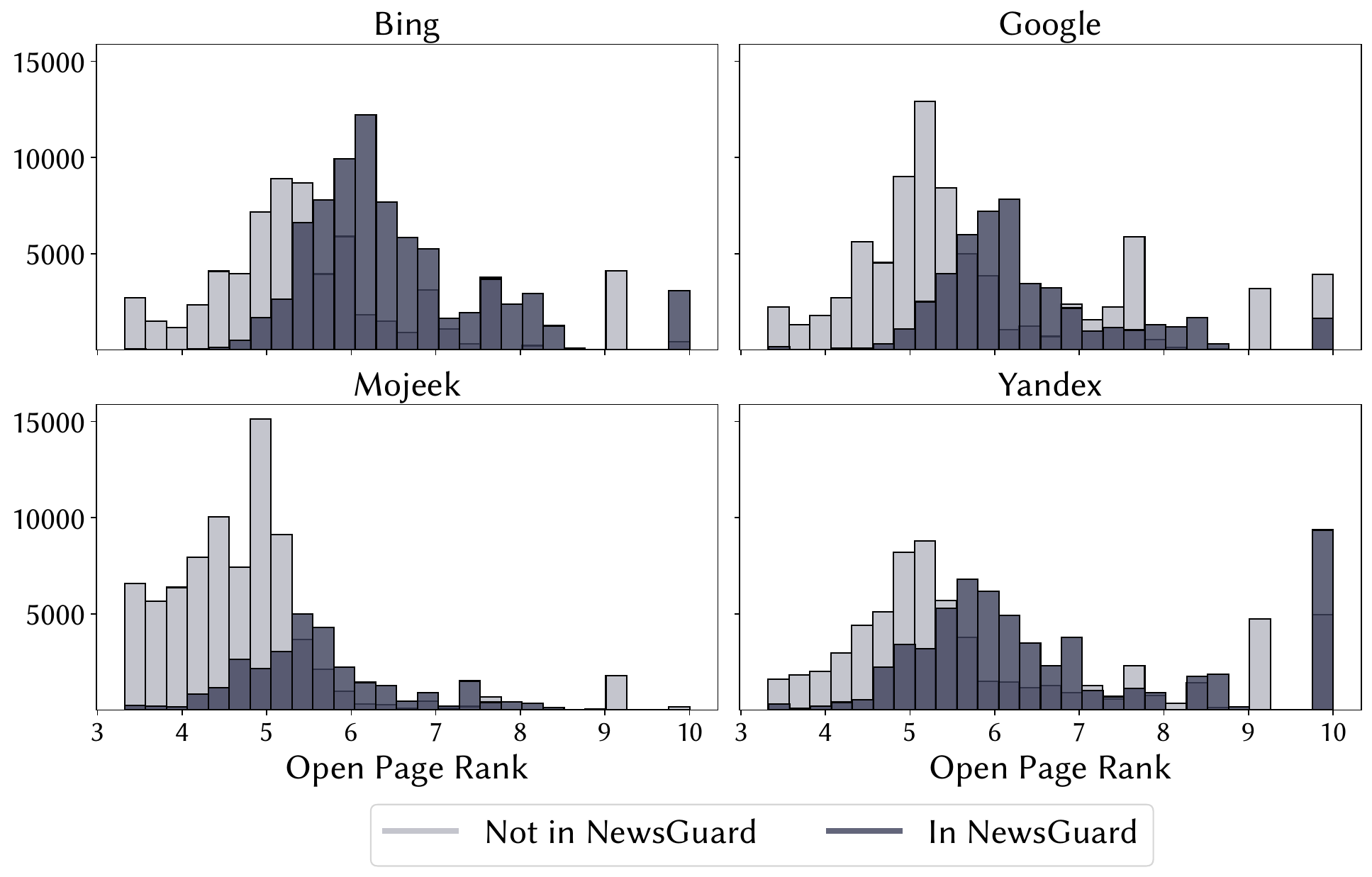}
  \captionof{figure}{Open PageRank scores of search results with and without NewsGuard metadata. An Open PageRank of 10 represents the most popular and authoritative web-domains.}
  \label{fig:NG_OPR}
\end{minipage} \hfill
\begin{minipage}[t]{0.35\linewidth}
  \centering
  \includegraphics[height=3.7cm, trim=5 -48 5 5, clip]{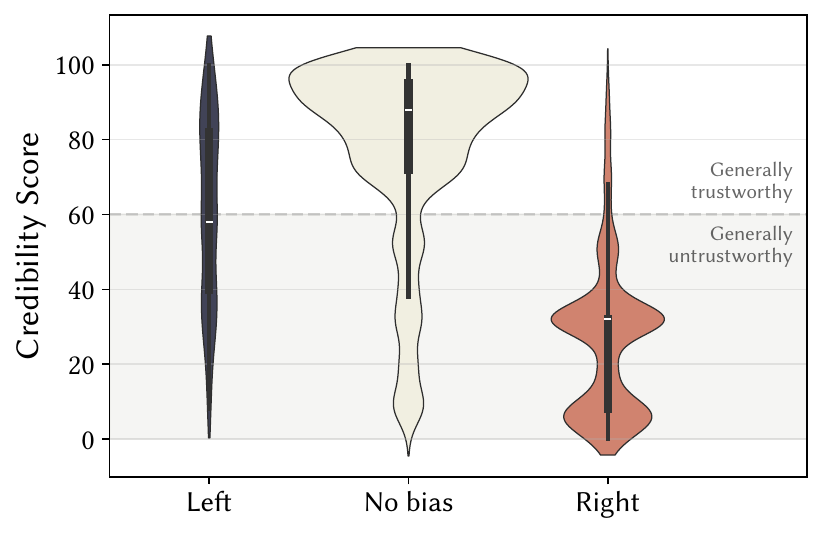}
  \captionof{figure}{Credibility per political orientation label for full NewsGuard database.}
  \label{fig:violin_NG}
\end{minipage}
\end{figure}

As is evident in Tables \ref{table:coverage_per_Country} and \ref{table:coverage_per_SE}, most search results and domains do not have NewsGuard metadata. This is not surprising, since our queries may result in obscure sites that have been of lower priority for NewsGuard to evaluate. To validate this, we examined the Open PageRank\footnote{Open PageRank is based on Google's PageRank algorithm, which is no longer publicly visible since 2016. It is apparently still in use as a minor part of their ranking algorithm \cite{domcopOpenPR,semrushGooglePageRank}.} of the websites in our dataset with and without metadata. Websites without NewsGuard metadata generally have a lower Open PageRank \cite{domcopOpenPR}, meaning they are ranked lower by search engines and therefore less likely to be shown (see Fig \ref{fig:NG_OPR}). 

There is also a lower coverage of unique domains compared to search results. This indicates that there are a lot of "long-tail" web-domains in our data. Domains with NewsGuard metadata are more popular and therefore appear more frequently, boosting the coverage of search results. Mojeek's search results (and unique domains) are covered the least by NewsGuard, which reflects the search engine's commitment to providing more diverse search results, even if the sites are not necessarily deemed authoritative \cite{mojeekMojeekRanking}. Surprisingly Google has lower NewsGuard coverage compared to Bing or Yandex. %
Overall, given the extraordinary number of websites in existence, NewsGuard's coverage, though not complete, is more than adequate. 

\subsection{Distribution of credibility scores per political orientation in NewsGuard}

As a point of comparison, we provide an overview of the credibility score distribution for left-, right-, and no political orientations. As can be seen, the credibility scores range the entire spectrum for all orientations, but right-leaning content tends to also score lower in terms of credibility. 
While it is tempting to relegate the difference in credibility distributions per political leaning to labelling bias, NewsGuard is currently one of the most extensive, detailed, and high-quality databases covering online sources. As such, the database presents one of the better existing approximations of the online sphere.

\subsection{Absolute counts per political orientation label per credibility score range}\label{app:counts_per_se_label_and_cred_range}

To provide thorough insight into the dataset, we also provide the counts of the political orientation labels per credibility score range, for each search engine. These results combine all data across all query topic categories, and all countries.

\begin{table}[ht]
\centering
\footnotesize
\caption{Counts per political orientation label and credibility score range for Bing.}
    \begin{tabular}{p{0.1\textwidth}  R{0.1\textwidth}  R{0.1\textwidth}  R{0.1\textwidth}  R{0.1\textwidth}  R{0.1\textwidth}}
        \toprule
        \textbf{Bing} & 0-20 & 20-40 & 40-60 & 60-80 & 80-100 \\
        \midrule
        Left & 0 & 0 & 11 & 1201 & 8530 \\
        No Bias & 17 & 13 & 278 & 2619 & 58619 \\
        Right & 192 & 128 & 223 & 387 & 1263 \\
        \bottomrule
    \end{tabular}
\end{table}

\begin{table}[ht]
    \centering
    \footnotesize
    \caption{Counts per political orientation label and credibility score range for Google.}
    \begin{tabular}{p{0.1\textwidth}  R{0.1\textwidth}  R{0.1\textwidth}  R{0.1\textwidth}  R{0.1\textwidth}  R{0.1\textwidth}}
        \toprule
        \textbf{Google} & 0-20 & 20-40 & 40-60 & 60-80 & 80-100 \\
        \midrule
        Left & 28 & 2 & 139 & 780 & 6223 \\
        No Bias & 6 & 60 & 171 & 1729 & 31755 \\
        Right & 213 & 145 & 357 & 768 & 1072 \\
        \bottomrule
    \end{tabular}
\end{table}

\begin{table}[ht]
    \centering
    \footnotesize
    \caption{Counts per political orientation label and credibility score range for Mojeek.}
    \begin{tabular}{p{0.1\textwidth}  R{0.1\textwidth}  R{0.1\textwidth}  R{0.1\textwidth}  R{0.1\textwidth}  R{0.1\textwidth}}
        \toprule
        \textbf{Mojeek} & 0-20 & 20-40 & 40-60 & 60-80 & 80-100 \\
        \midrule
        Left & 91 & 94 & 20 & 1482 & 3562 \\
        No Bias & 244 & 279 & 361 & 1273 & 10887 \\
        Right & 3519 & 2484 & 1092 & 1326 & 1699 \\
        \bottomrule
    \end{tabular}
\end{table}

\begin{table}[H]
    \centering
    \footnotesize
    \caption{Counts per political orientation label and credibility score range for Yandex.}
    \begin{tabular}{p{0.1\textwidth}  R{0.1\textwidth}  R{0.1\textwidth}  R{0.1\textwidth}  R{0.1\textwidth}  R{0.1\textwidth}}
        \toprule
        \textbf{Yandex} & 0-20 & 20-40 & 40-60 & 60-80 & 80-100 \\
        \midrule
        Left & 13 & 46 & 450 & 2100 & 6677 \\
        No Bias & 687 & 931 & 353 & 2389 & 22264 \\
        Right & 3364 & 1787 & 1302 & 2727 & 1830 \\
        \bottomrule
    \end{tabular}
\end{table}

\end{document}